# Thermodynamic extension of density-functional theory.
# I. Basic Massieu function, its Legendre and Massieu-Planck transforms for equilibrium state in terms of density matrix

Robert Balawender[1,2,a] and Andrzej Holas[1]

1) Institute of Physical Chemistry of Polish Academy of Sciences, Kasprzaka 44/52, PL-01-224 Warsaw, Poland
2) Eenheid Algemene Chemie (ALGC), Vrije Universiteit Brussel, Pleinlaan 2, B-1050 Brussels, Belgium

A general formulation of the equilibrium state of a many-electron system in terms of a (mixed-state, ensemble) density matrix operator in the Fock space, based on the maximum entropy principle, is introduced. Various state functions/functionals are defined and investigated: the basic Massieu function for fully open thermodynamic system, the effective action function for the fully closed (isolated) system, and a series of Legendre transforms for partially open/closed ones — the Massieu functions. Convexity and concavity properties of these functions are determined, their first and second derivatives with respect to all arguments are obtained. Other state functions — the Gibbs-Helmholtz functions — are obtained from previous ones as their Massieu-Planck transforms, i.e. by multiplying them by minus temperature and by specific transformation of arguments (which involves the temperature). Such functions are closer to traditional (Gibbs, Helmholtz) thermodynamic potentials. However, the first and second derivatives of these functions represent more complicated expressions than derivatives of the Massieu functions. All introduced state functions are suitable for application to various extensions of the density functional theory (DFT), both at finite temperature and at zero-temperature limit. Derivatives of state functions are essential for determining the chemical reactivity and other descriptors of the conceptual DFT extensions.

---

[a] Author to whom correspondence should be addressed, rbalawender@ichf.edu.pl

## I. INTRODUCTION

Density-functional theory (DFT) is a rigorous approach for describing the ground state of any electronic system.[1-5] The success and popularity of DFT are based on the fact that the fundamental variable — the electron density — is an observable, and also that it is a much simpler object than the many-body wave or Green's function. Simple approximations of the exchange-correlation energy functional, used by implementations of DFT, perform remarkably well for a wide range of problems in chemistry and physics[6-8], particularly for predictions of the structure and thermodynamic properties of molecules and solids. Despite successes of DFT, its applications still suffer from errors that cause qualitative failures in some predicted properties. A systematic approach to constructing universally applicable functionals is a hard problem and has remained elusive. A possible step forward is to recover a balance between the qualitative (conceptual) and the quantitative (computational) branches of DFT, and to have a deeper look at violations, by currently used functionals, of the exact conditions that should be satisfied by DFT.[9-17] For example, the delocalization error originates from the violation of linearity of the energy as a function of fractional charges,[18-22] the static correlation error emerges from the violation of constancy of the energy as a function of fractional spins.[20,23,24] Both are important to assess errors of functionals.[25] Distressingly, the status of nonintegral electron number, fractional spins and fractional occupations in the DFT is problematic.[9,26-33] The answer to the question how system properties depend on the electron number is crucial. Conceptual DFT offers a perspective on the interpretation and prediction of experimental and theoretical reactivity data based on a series of response functions to perturbations in the number of electrons and/or external potential. This approach enables a sharp definition and then computation, from first principles, of a series of well-known but sometimes vaguely defined chemical concepts such as electronegativity, hardness and Fukui function. Identification of chemical concepts and principles in terms of derivatives with respect to the electron density (local chemical reactivity descriptors) or with respect to the total electron number (global properties)[34-42] requires an extension of the density domain to ensemble density which integrates to a real positive electron number.

There are two main approaches to the introduction of a nonintegral (fractional) particle number into DFT[32,33,43,44]. First approach applies directly a smooth extension of the expression for the electron density and the kinetic energy by introducing fractional occupations of the involved orbitals, and simultaneously applies such density as an argument of the approximate functionals (like LDA or GGA ones), that were originally derived as approximations for the integral-electron-number systems, but remain mathematically meaningful for arbitrary density. In the second, thermodynamic approach, the extension from an integral to fractional electron number is realized through a grand canonical ensemble of Mermin's finite-temperature DFT[26,45-48] at zero-temperature limit. Besides these two approaches, one can also work out the connection with the correct integral-$N$ functionals by applying suitable interpolation methods,[49-51] e.g. a quadratic approximation of $E(N)$. While various ways for an extension from integral- to fractional-number approaches may be mathematically acceptable (correct), nevertheless important physical arguments exist that remove arbitrariness. Based on the axiomatic formulations (formal mathematical properties) of the Hohenberg-Kohn functional, only such functional that satisfies the variational principle for the electronic energy as a functional of the density, and is continuous, convex and size consistent can be accepted as the exact functional.[9,11,52] The functional defined with the thermodynamic approach satisfies these conditions. Moreover, this functional is identical to the functional obtained using the Legendre transformation and density-matrix constrained search techniques. No functional consistent with the variational principle can reach a lower minimum than this functional.[52] Another argument is that only the thermodynamic approach is the proper basis for investigation of open systems and is capable of tackling all issues in the chemical reactivity.[53]

The finite-temperature formalism of DFT was formally set up long ago by Mermin.[45] Only a few generalizations have been reported so far.[47,54-59] We wish to extend the density domain of DFT to the spin-polarized density, which can integrate to a real positive number of particles and a real number of spins. Before developing this formalism (which will be done in the next papers in the series), we will concentrate in this paper – the first in the series – on properties of the equilibrium state of the many-electron system expressed in terms of the density matrix operator in the Fock space. Results of the present paper may also provide foundations for other extensions of DFT, like the current-density functional theory, the (one-particle) density-matrix functional theory, the pair-density functional theory, etc.

In Sec. II, we provide the entropic formulation of the equilibrium state using the maximum entropy principle. Based on the effective action formalism[60,61] and Jensen inequality, convexity and concavity properties of the basic Massieu function[62] and its functional Legendre transform – the effective action function – are established. Although in the general thermostatics the exponential function appearing in the Boltzmann-Gibbs distributions can be replaced by some other increasing function,[63,64] only the well-known Boltzmann-Gibbs-Shannon entropy is considered here. The results derived in Sec. II which characterize a fully open and fully closed system are translated in Sec. III to partially open/closed systems by applying the Legendre transformation mechanism. Properties of all these functions — named the Massieu functions — are discussed, in particular, their convexity and concavity. The relation between the second-derivative matrix and the covariance matrix is derived. In Sec. IV, the relations between equilibrium state functions defined by the above formal mathematical procedure (Massieu functions) and by those defined in analogy with the traditional (e.g. Gibbs and Helmholtz) potentials (here Gibbs-Helmholtz functions) are investigated using the Massieu-Planck transformation.[65] Table I displays relations between all mentioned state functions. The expression for the second derivative matrix of the Gibbs-Helmholtz functions in terms of the second-derivative matrix of the corresponding Massieu functions makes a link between a real thermodynamic transformation (easy controlled in the laboratory) and that due to imposing constraints that appear more natural in the maximum entropy principle. The preeminence of Massieu functions over the Gibbs-Helmholtz functions in a compact formulation of a generalized statistical mechanics and thermodynamics,[66] and in undergraduate instruction[67-70] is evident. However the standard reference books on thermodynamics usually do not give enough information about these functions; the axiomatic thermodynamics introduced by Callen[71], is, to our best knowledge, an exception. Advantages of the Massieu functions over the Gibbs-Helmholtz functions in analyzing the properties of the equilibrium state become apparent when we compare complexity of the expressions for their second derivatives. Sec. V is devoted to conclusions. In Appendix A, the derivatives of the Massieu functions and the Gibbs-Helmholtz functions are obtained using the matrix notation. This Appendix may be useful also for a "pedagogical" reason. In Appendix B the notion of positive (negative) definiteness of a matrix is recalled. In Appendix C the definition and properties of Legendre transforms are recalled. Appendix D provides a collection of identities satisfied by the vector functions of thermodynamic variable transformations.

*Note:* Atomic units are used throughout the paper. The abbreviation "w.r.t." means "with respect to", "fn." means "function". "Tr" denotes the trace operation in the Fock space. "$\Sigma$" means a summation over all indices, however if the summation index (variable) is continuous, "$\Sigma$" means an integration. See Appendix A for details concerning the vector and matrix notations.

# II. ENTROPIC FORMULATION OF THE EQUILIBRIUM STATE USING THE MAXIMUM ENTROPY PRINCIPLE. CHARACTERISTIC STATE FUNCTIONS.

The mathematical description of a molecule as a many-electron system of interest is based on a state-vector linear space; generally, it will be the Fock space in the second quantization formulation, and operators in this space written in terms of the field operators $\hat{\psi}(\mathbf{r},\sigma)$ and $\hat{\psi}^{\dagger}(\mathbf{r},\sigma)$. Microscopically (quantum mechanically) such a system is defined by its time-independent non-relativistic Hamiltonian operator $\hat{H}$, whereas in the macroscopic domain of the thermodynamics it is specified by a set of observables $\{\hat{O}_j\}$ — the generating operators. The corresponding system will be named the $\{\hat{O}_j\}$ thermodynamic system for this molecule. All generating operators are assumed to be time-independent, hermitian, local (**r**-dependent) or global (**r**-independent), and mutually commuting, $[\hat{O}_i, \hat{O}_j]_- = 0$.

The thermodynamic system is characterized by a density matrix (DM) operator $\hat{\Gamma}$ (hermitian, positive semidefinite and the unit-trace one). For a general mixed state, $\hat{\Gamma}$ can be expanded in terms of its eigenvalues and eigenstates

$$\hat{\Gamma} = \hat{\Gamma}\big[\{g_K\},\{\Psi_K\}\big] = \sum_K |\Psi_K\rangle g_K \langle\Psi_K|, \quad \langle\Psi_K|\Psi_L\rangle = \delta_{KL}, \quad 0 \leq g_K, \quad \mathrm{Tr}\,\hat{\Gamma} = \sum_K g_K = 1, \quad (1)$$

where the set $\{|\Psi_K\rangle\}$ of Fock-space vectors is complete, $\sum_K |\Psi_K\rangle\langle\Psi_K| = \hat{1}$. The expectation (average) value of any operator $\hat{O}_j$ is defined in terms of $\hat{\Gamma}$ as

$$\langle\hat{O}_j\rangle_\Gamma = \mathrm{Tr}\,\hat{\Gamma}\hat{O}_j = \sum_K g_K \langle\Psi_K|\hat{O}_j|\Psi_K\rangle. \tag{2}$$

The expansion (1) was applied for the second form. Along with averages, the first-order covariance matrix of expectation values may also be of interest. Its elements are defined as

$$\left\langle\left\langle\hat{O}_i,\hat{O}_j\right\rangle\right\rangle_\Gamma = \left\langle\hat{O}_i\hat{O}_j\right\rangle_\Gamma - \left\langle\hat{O}_i\right\rangle_\Gamma\left\langle\hat{O}_j\right\rangle_\Gamma = \left\langle\left(\hat{O}_i - \left\langle\hat{O}_i\right\rangle_\Gamma\right)\left(\hat{O}_j - \left\langle\hat{O}_j\right\rangle_\Gamma\right)\right\rangle_\Gamma. \qquad (3)$$

Diagonal elements correspond to the variance of the expectation values, where $\left(\hat{O}_i - \left\langle\hat{O}_i\right\rangle_\Gamma\right)$ is the fluctuation of the expectation value.

All sets of possible average values of the operators of the $\{\hat{O}_j\}$ system define a space of so-called thermodynamic states. To each *average-value* variable corresponds a *conjugate* variable of the environment — the *external source*. Starting from an all-average state (its independent variables are only average values), one can get different states (with different sets of independent variables) by replacing some average values by their conjugates — the corresponding external sources. The thermodynamic system can be classified according to the state variables used to describe it: the fully closed (isolated) one means that only average values are used as the state variables; the fully open one means that only external sources are used as the state variables; and various intermediate, partially open/closed systems can be considered too. As a thermodynamic process, we consider a sequence of state changes defined in the state space of a set of average values and/or external sources. We will consider only reversible processes, which must consist of equilibrium states only; the existence of equilibrium state is taken as a fundamental fact of experience. So, there are various types of the equilibrium state: fully closed, fully open, and intermediate types. These types will be named also thermodynamic ensembles, see Sec. III

The *equilibrium state* is completely decrypted by the equilibrium density matrix (eq-DM) operator $\hat{\Gamma}_{eq}$. For its determination we use the principle of the maximum for the informational entropy, based on the phenomenological description.[72,73] In order to find the equilibrium state for fully closed systems, the maximization is constrained by the demand that the expectation values of the observables $\{\hat{O}_j\}$ have chosen values $\{o_j\}$. These values represent the independent variables characterizing the state of the closed system. Irrespectively of the specific form of the entropy $S^{\text{ent}}[\hat{\Gamma}]$, the realization of the principle of the conditional maximum reduces to the variational procedure for finding the unconditional maximum of the Lagrange function (fn.)

$$\Lambda\left[\hat{\Gamma}, \hat{O}[\{\alpha_j\},\{\hat{O}_j\}]\right] = S^{\text{ent}}[\hat{\Gamma}] - \sum_j \alpha_j \langle \hat{O}_j \rangle_\Gamma = S^{\text{ent}}[\hat{\Gamma}] + \text{Tr}\,\hat{\Gamma}\hat{O} \qquad (4)$$

($\{\alpha_j\}$ is a set of Lagrange multipliers), where the "general conditions operator" is

$$\hat{O} = \hat{O}[\{\alpha_i\},\{\hat{O}_i\}] = -\sum_j \alpha_j \hat{O}_j. \qquad (5)$$

When $\hat{\Gamma}_{\text{eq}}[\hat{O}]$ is determined from maximization of the Lagrange fn., Eq.(4), the required values of the parameters $\{o_j\}$ of the closed system can be attained by choosing such values of multipliers $\{\alpha_j\}$ that the requirements

$$\langle \hat{O}_j \rangle_{\hat{\Gamma}_{\text{eq}}} \equiv \text{Tr}\,\hat{\Gamma}_{\text{eq}}\left[\hat{O}[\{\alpha_i\},\{\hat{O}_i\}]\right]\hat{O}_j = o_j. \qquad (6)$$

are satisfied. But before imposing the requirements (6), we can consider the equilibrium state described by $\{\alpha_j\}$ as independent parameters. By definition, it is termed the equilibrium state of a fully open system, while the parameters (Lagrange multipliers) $\{\alpha_j\}$ represent the external sources mentioned earlier.

To make our paper an easier read, we are going to indicate in various places the exemplary case which covers statistical mechanics problems (ensembles, thermodynamic functions etc.) considered in the Parr-Yang book.[1] Here, the thermodynamic system of this case is specified by $\{\hat{O}_1, \hat{O}_2\} = \{\hat{H}, \hat{N}\}$, where $\hat{N}$ is the particle-number operator.

Using for the entropy the Boltzmann-Gibbs-Shannon (BGS) expression — the expectation value of the natural logarithm of $\hat{\Gamma}$ (with a minus sign; the Boltzmann constant is omitted because we will use temperature expressed in energy unit)

$$S^{\text{ent}}\left[\hat{\Gamma}\right] = S^{\text{BGS}}\left[\hat{\Gamma}\right] \equiv -\text{Tr}\,\hat{\Gamma}\ln\hat{\Gamma}, \qquad (7)$$

from the usual Euler approach — the requirement $\delta\Lambda = 0$ — the well-known canonical distribution, adequately describing the physical system, is obtained

$$\hat{\Gamma}_{\text{eq}}\left[\hat{O}\right] = \exp\left(\hat{O}\right)\big/\Xi\left[\hat{O}\right], \qquad (8)$$

where

$$\Xi\left[\hat{O}\right] = \text{Tr}\exp\left(\hat{O}\right) \qquad (9)$$

is the *partition fn*. The result of maximization of the Lagrange fn., Eq.(4), will be named the *basic Massieu fn.*:

$$\Lambda_{\text{eq}}\left[\hat{O}\right] = \underset{\{\hat{\Gamma}:\text{Tr}\,\hat{\Gamma}=1\}}{\text{Max}}\Lambda\left[\hat{\Gamma},\hat{O}\right] = \Lambda\left[\hat{\Gamma}_{\text{eq}},\hat{O}\right] = S^{\text{BGS}}\left[\hat{\Gamma}_{\text{eq}}\right] - \sum_j \alpha_j \text{Tr}\hat{\Gamma}_{\text{eq}}\hat{O}_j, \qquad (10)$$

a dimensionless quantity, while the maximizer $\hat{\Gamma}_{\text{eq}} = \hat{\Gamma}_{\text{eq}}\left[\hat{O}\right]$ — the *eq-DM operator*, Eq.(8). When the label $j$ of an independent argument variable $\alpha_j$ of $\Lambda_{\text{eq}}$ is a continuous parameter, e.g., the position vector $\mathbf{r}$, $\alpha_j \Rightarrow \alpha(\mathbf{r})$, the fn. $\Lambda_{\text{eq}}$ becomes a functional of this variable. It is easy to check that $\Lambda_{\text{eq}}$ can be expressed in terms of $\Xi$ as (see Eqs.(4) and (8))

$$\Lambda_{\text{eq}}\left[\hat{O}\right] = \Lambda\left[\hat{\Gamma}_{\text{eq}}\left[\hat{O}\right],\hat{O}\right] = \ln\Xi\left[\hat{O}\right]. \qquad (11)$$

This form (11) of $\Lambda_{\text{eq}}$ allows for investigation of its properties (like convexity) using the effective action formalism.

The above expressions for $\hat{\Gamma}_{\text{eq}}\left[\hat{O}\right]$ and $\Lambda_{\text{eq}}\left[\hat{O}\right]$ are meaningful provided the partition fn., Eq.(9) with Eq.(5), attains a finite value:

$$\Xi\left[\hat{O}\right] = \sum_K \left\langle \Psi_K^{\text{cev}} \left| \exp\left(\hat{O}\right) \right| \Psi_K^{\text{cev}} \right\rangle = \sum_K \exp\left(O^K\right) < +\infty. \qquad (12)$$

Here, the state vectors $|\Psi_K^{\text{cev}}\rangle$ are, due to assumed $[\hat{O}_i, \hat{O}_j]_- = 0$, the common eigenvectors (cev) of all operators $\hat{O}_j$ chosen for the definition of the thermodynamic system:

$$\hat{O}_j |\Psi_K^{\text{cev}}\rangle = o_j^K |\Psi_K^{\text{cev}}\rangle. \tag{13}$$

The set $\{|\Psi_K^{\text{cev}}\rangle\}$ is complete, its elements are normalized and mutually orthogonal. It will be convenient now to apply the convention described in Appendix A and, instead of using $M$-element sets like $\{\alpha_j\}_{j=1}^M, \{o_j\}, \{\hat{O}_j\}$, to introduce vectors $\boldsymbol{\alpha}_\mathsf{F} = (\alpha_1\ \alpha_2\ ...\ \alpha_M)^\mathrm{T}$, $\boldsymbol{o}_\mathsf{F} = (o_1\ o_2\ ...\ o_M)^\mathrm{T}$, $\hat{\boldsymbol{O}}_\mathsf{F} = (\hat{O}_1\ \hat{O}_2\ ...\hat{O}_M)^\mathrm{T}$. For exemplary case we have $M = 2$, $\mathsf{F} = \{1, 2\}$, $\boldsymbol{\alpha}_\mathsf{F} = (\alpha_1, \alpha_2)^\mathrm{T}$, $\hat{\boldsymbol{O}}_\mathsf{F} = (\hat{H}, \hat{N})^\mathrm{T}$. The conditions operator, Eq.(5), represents a scalar product $\hat{O} = -\hat{\boldsymbol{O}}_\mathsf{F}^\mathrm{T} \boldsymbol{\alpha}_\mathsf{F} = -\boldsymbol{\alpha}_\mathsf{F}^\mathrm{T} \hat{\boldsymbol{O}}_\mathsf{F}$. Therefore its eigenvalue at the eigenvector $|\Psi_K^{\text{cev}}\rangle$ occurring in Eq.(12), is

$$O^K = -\boldsymbol{\alpha}_\mathsf{F}^\mathrm{T} \boldsymbol{o}_\mathsf{F}^K. \tag{14}$$

The restriction written in Eq.(12) with (14) may be violated in two cases: (i) the eigenvalue spectra $\boldsymbol{o}_\mathsf{F}^K = \boldsymbol{o}_\mathsf{F}^K[\hat{\boldsymbol{O}}_\mathsf{F}]$ happen to be such that the summation over $K$ is divergent for every point $\boldsymbol{\alpha}_\mathsf{F}$ from the external-sources space, (ii) while the summation is convergent for some $\boldsymbol{\alpha}_\mathsf{F}$, it may be divergent for another $\boldsymbol{\alpha}_\mathsf{F}$. Thus, not all sets of observables are suitable for the specification of the thermodynamic system. The *valid* set is such that the restriction, Eq.(12) with (14), is satisfied for at least one point $\boldsymbol{\alpha}_\mathsf{F}$. For example, considering the set $\hat{\boldsymbol{O}}_\mathsf{F}$ of observables having positive, extending to infinity spectra $\left\{\operatorname{Inf}_K o_j^K = o_j^{\min} > 0,\ \operatorname{Sup}_K o_j^K = +\infty\right\}_{j=1}^M$, the point $\{\alpha_j = 1/o_j^{\min}\}_{j=1}^M$ seems to be the mentioned point suitable for checking the validity. All further considerations in the present paper will be confined to valid only sets of observables.

For the valid set $\hat{O}_F$ we define the domain $\mathcal{A} = \mathcal{A}[\hat{O}_F]$ of the acceptable external-sources points by

$$\sum_K \exp(-\alpha_F o_F^K) < +\infty, \qquad \forall \alpha_F \in \mathcal{A}[\hat{O}_F]. \tag{15}$$

For example, the point $\alpha_F = (0,0,...,0) \notin \mathcal{A}[\hat{O}_F]$ because $\sum_K 1 = +\infty$ for infinite set of state vectors. Finally, the codomain of acceptable expectation-values points $\mathcal{O}[\hat{O}_F]$ is defined by the mapping (see Eq.(6))

$$\alpha_F \mapsto o_F = o_F^{eq}[\alpha_F, \hat{O}_F] \equiv \mathrm{Tr}\,\hat{\Gamma}_{eq}[\hat{O}[\alpha_F, \hat{O}_F]]\hat{O}_F \tag{16}$$

namely

$$\mathcal{O}[\hat{O}_F] = o_F^{eq}[\mathcal{A}[\hat{O}_F], \hat{O}_F]. \tag{17}$$

It is important that $\hat{\Gamma}_{eq}$ given in Eq.(8) not only satisfies $\delta\Lambda = 0$ but also it is the true maximizer, i.e. that the variational principle holds: $\Lambda[\hat{\Gamma}, \hat{O}] \leq \Lambda[\hat{\Gamma}_{eq}[\hat{O}], \hat{O}]$. The proof is based on the Jensen's inequality for convex functions of an operator. For the proof, it may be convenient to have $\hat{\Gamma}$ in the form (1). Then, let us perform the maximization, Eq.(10), first w.r.t. the weights $g_K$ of the expanded $\hat{\Gamma} = \hat{\Gamma}[\{g_K\}, \{\Psi_K\}]$, Eq.(1), at fixed $\{\Psi_K\}$. The optimum weights $\{p_K\}$ are found in the form of the Boltzmann-Gibbs distribution, the canonical distribution $p_K = p_K[\{\Psi_L\}, \hat{O}] = \exp\langle\Psi_K|\hat{O}|\Psi_K\rangle / \sum_L \exp\langle\Psi_L|\hat{O}|\Psi_L\rangle$. The next maximization w.r.t. $\{\Psi_K\}$ leads to $\{\Psi_K^{cev}\}$, which are the eigenfunctions of $\hat{\Gamma}_{eq}$, and due to Eq.(8), simultaneously the eigenfunctions of $\hat{O}$, Eqs.(13),(14). The obtained eq-DM is $\hat{\Gamma}_{eq}[\hat{O}] = \sum_K p_K^{eq} |\Psi_K^{cev}\rangle\langle\Psi_K^{cev}|$ with $p_K^{eq} = p_K[\{\Psi_L^{cev}\}, \hat{O}]$. The following inequalities (based on the Gibbs inequality and the Jensen's inequality) hold

$$\Lambda\left[\hat{\Gamma}\left[\{g_K\},\{\Psi_K\}\right]\right] \leq \Lambda\left[\hat{\Gamma}\left[\{p_K\},\{\Psi_K\}\right]\right] \leq \Lambda\left[\hat{\Gamma}\left[\{p_K^{eq}\},\{\Psi_K^{cev}\}\right]\right] = \Lambda\left[\hat{\Gamma}_{eq}\right] = \Lambda_{eq} \qquad (18)$$

(dependence on $\hat{O}\left[\alpha_F,\hat{O}_F\right]$ is suppressed here for brevity). The detailed proof of these inequalities can be performed analogously as in Ref.1, pp.64-65, where it is done on the example of $\hat{O} = -\beta\left(\hat{H} - \mu\hat{N}\right)$. It corresponds to the exemplary case with $\alpha_1 = \beta$, $\alpha_2 = -\beta\mu$. The fn. $\Lambda\left[\hat{\Gamma}, -\alpha_1\hat{H} - \alpha_2\hat{N}\right]$ here, Eq.(4), is the same as $-\beta\Omega\left[\hat{\Gamma},\beta,\mu\right] = -\alpha_1\Omega\left[\hat{\Gamma},\alpha_1,-\alpha_1^{-1}\alpha_2\right]$ where $\Omega$ is the grand potential of Ref.1, $\Omega\left[\hat{\Gamma},\beta,\mu\right] = \text{Tr}\,\hat{\Gamma}\left(\beta^{-1}\ln\hat{\Gamma} + \hat{H} - \mu\hat{N}\right)$. At equilibrium analogous relations hold between $\Lambda_{eq}$ and $\Omega^0$.

The basic Massieu fn. has the generating property for expectation values:

$$\left(\frac{\partial\Lambda_{eq}\left[\alpha_F,\hat{O}_F\right]}{\partial\alpha_F}\right) = -\text{Tr}\,\hat{\Gamma}_{eq}\left[\alpha_F,\hat{O}_F\right]\hat{O}_F = -\left\langle\hat{O}_F\right\rangle_{\Gamma_{eq}}, \qquad (19)$$

where the new notation $\Lambda_{eq}\left[\alpha_F,\hat{O}_F\right]$ means previous $\Lambda_{eq}\left[\hat{O}\left[\alpha_F,\hat{O}_F\right]\right]$ (similar convention will be applied to the arguments of $\hat{\Gamma}_{eq}$). This generating property can be easily shown by differentiation w.r.t. $\alpha_i$ of Eq.(11) with inserted Eq.(5). The generating property of an open system, Eq.(19), can be used now to obtain the properties of the closed system. The vector fn. $\alpha_F^{eq}\left[o_F,\hat{O}_F\right]$ – the equilibrium multipliers guaranteeing satisfaction of the constraints (6) – can be determined from Eq.(19) rewritten in a form of a vector equation

$$\left(\frac{\partial\Lambda_{eq}\left[\alpha_F,\hat{O}_F\right]}{\partial\alpha_F}\right)^T = -o_F, \qquad (20)$$

as its solution w.r.t. $\alpha_F$ at given $o_F$, $\hat{O}_F$ (the gradient of a fn. is a row vector, see Appendix A). It means that the identity $\text{Tr}\,\hat{\Gamma}_{eq}\left[\alpha_F^{eq}\left[o_F,\hat{O}_F\right],\hat{O}_F\right]\hat{O}_F = o_F$ should be satisfied at each point $o_F \in \mathcal{O}$. For the exemplary case, $o_F = (E, N)^T$.

The problem of existence and uniqueness of the solution of this vector equation needs special attention. Eq.(20) can be viewed as the direct map $\alpha_F \mapsto o_F$ in terms of $\Lambda_{eq}\left[\alpha_F,\hat{O}_F\right]$, or equivalently, of $\hat{\Gamma}_{eq}\left[\alpha_F,\hat{O}_F\right]$, namely

$$\alpha_F \mapsto o_F = o_F^{eq}\left[\alpha_F,\hat{O}_F\right] \equiv \text{Tr}\,\hat{\Gamma}_{eq}\left[\alpha_F,\hat{O}_F\right]\hat{O}_F = -\left(\frac{\partial \Lambda_{eq}\left[\alpha_F,\hat{O}_F\right]}{\partial \alpha_F}\right)^T \equiv -\Lambda_F\left[\alpha_F,\hat{O}_F\right] \quad (21)$$

(obtained earlier in Eq.(16) in terms of $\hat{\Gamma}_{eq}$ only). In the last equality the notation $\Lambda_F$ of Eq.(A2) for the gradient was applied together with suppression of the subscript 'eq'.

Based on the property (proven in Appendix A of Ref. [61]) of Hermitian operators $\hat{A}$ and $\hat{B}$ that the relation $\text{Tr}\exp(\varepsilon\hat{A} + (1-\varepsilon)\hat{B}) \leq (\text{Tr}\exp\hat{A})^\varepsilon (\text{Tr}\exp\hat{B})^{(1-\varepsilon)}$ is satisfied for $0 < \varepsilon < 1$ (equality holds if and only if operators $\hat{A}$ and $\hat{B}$ differ by an ordinary number), the *strict convexity* of $\Lambda_{eq}\left[\alpha_F,\hat{O}_F\right]$ as a function of $\alpha_F$ can be proven.[61] It means that inequality

$$\Lambda_{eq}\left[\varepsilon\alpha_F^{(0)} + (1-\varepsilon)\alpha_F^{(1)},\hat{O}_F\right] < \varepsilon\Lambda_{eq}\left[\alpha_F^{(0)},\hat{O}_F\right] + (1-\varepsilon)\Lambda_{eq}\left[\alpha_F^{(1)},\hat{O}_F\right] \quad (22)$$

holds for $0 < \varepsilon < 1$ and for $\alpha_F^{(0)} \neq \alpha_F^{(1)}$. This property is of fundamental importance: based on it, the one-to-one map $\alpha_F \leftrightarrow o_F$ can be proven.[61] Therefore, the solution $\alpha_F = \alpha_F^{eq}\left[o_F,\hat{O}_F\right]$ of Eq.(20) is unique, and it represents the map $o_F \mapsto \alpha_F$, reciprocal to that in Eq.(21).

Due to the generating property (20) of $\Lambda_{eq}$, the change of variables from $\alpha_F$ to $o_F$ can be accomplished via the Legendre transform $\Phi_{eq}\left[o_F, \hat{O}_F\right]$ of $\Lambda_{eq}$ (see Appendix C) known as the *effective action fn.*:

$$\Phi_{eq}\left[o_F, \hat{O}_F\right] \equiv \Lambda_{eq}\left[\alpha_F^{eq}\left[o_F, \hat{O}_F\right], \hat{O}_F\right] + o_F^T \alpha_F^{eq}\left[o_F, \hat{O}_F\right], \tag{23}$$

which is *strictly concave* fn. of the expectation values $o_F$.[61] From differentiation of Eq.(23) w.r.t. $o_F$ and after applying Eq.(20), one obtains the explicit map

$$o_F \mapsto \alpha_F = \alpha_F^{eq}\left[o_F, \hat{O}_F\right] \equiv \left(\frac{\partial \Phi_{eq}\left[o_F, \hat{O}_F\right]}{\partial o_F}\right)^T \equiv \Phi_F\left[o_F, \hat{O}_F\right], \tag{24}$$

showing the generating property of $\Phi_{eq}$. The mapping (24) defines also the relation between $\mathcal{O}\left[\hat{O}_F\right]$ and $\mathcal{A}\left[\hat{O}_F\right]$ as

$$\mathcal{A}\left[\hat{O}_F\right] = \alpha_F^{eq}\left[\mathcal{O}\left[\hat{O}_F\right], \hat{O}_F\right] \tag{25}$$

which is reciprocal to the relation in Eq.(17). Therefore, we observe the one-to-one correspondence between two domains $\mathcal{O}\left[\hat{O}_F\right]$ and $\mathcal{A}\left[\hat{O}_F\right]$. Since the vector of observable operators $\hat{O}_F$ is fixed, dependence on it will be often suppressed, e.g., $\hat{\Gamma}_{eq}\left[\alpha_F, \hat{O}_F\right] \Rightarrow \hat{\Gamma}_{eq}\left[\alpha_F\right]$. The proven maps $\alpha_F \leftrightarrow o_F$ yield identities (comp. Eqs.(C9) and (C8))

$$o_F^{eq}\left[\alpha_F^{eq}\left[o_F\right]\right] = o_F, \quad \text{i.e.,} \quad -\Lambda_F\left[\Phi_F\left[o_F\right]\right] = o_F, \tag{26}$$

$$\alpha_F^{eq}\left[o_F^{eq}\left[\alpha_F\right]\right] = \alpha_F, \quad \text{i.e.,} \quad \Phi_F\left[-\Lambda_F\left[\alpha_F\right]\right] = \alpha_F. \tag{27}$$

By expressing $\Lambda_{eq}$ in Eq.(23) in terms of the entropy, Eq.(10), we find that the effective action fn. is, in fact, just the entropy

$$\Phi_{eq}\left[o_F\right] = S^{BGS}\left[\hat{\Gamma}_{eq}^{av}\left[o_F\right]\right], \tag{28}$$

where $\hat{\Gamma}_{eq}^{av}[o_F] = \hat{\Gamma}_{eq}[\alpha_F^{eq}[o_F]]$ is the eq-DM operator expressed as a function of average values (av) $o_F$ (as independent variables).

Similarly as the Massieu fn. $\Lambda_{eq}$ in Eq.(11), the effective action fn. can be also obtained from appropriate partition fn.

$$\Phi_{eq}[o_F] = \ln \Xi[\hat{O}^{av}[o_F]], \quad (29)$$

using the specific conditions operator

$$\hat{O}^{av}[o_F] = -(\alpha_F^{eq}[o_F])^T (\hat{O}_F - o_F), \quad (30)$$

see Eqs.(8) and (9) with $\hat{O}$ replaced by $\hat{O}^{av}$.

As discussed in Appendix C, when $\Phi_{eq}$ is the Legendre transform of $\Lambda_{eq}$, Eq.(23), induced by the generating property of $\Lambda_{eq}$, Eq.(20), then $\Lambda_{eq}$ is the Legendre transform of $\Phi_{eq}$, induced by the generating property of $\Phi_{eq}$, Eq.(24), namely

$$\Lambda_{eq}[\alpha_F] = \Phi_{eq}[o_F^{eq}[\alpha_F]] - \alpha_F^T o_F^{eq}[\alpha_F] = S^{BGS}[\hat{\Gamma}_{eq}[\alpha_F]] - \alpha_F^T o_F^{eq}[\alpha_F] \quad (31)$$

(the relations (27) and (28) were used to obtain the entropy representation). Thus the fns. $\Lambda_{eq}$ and $\Phi_{eq}$ are Legendre-transformation *equivalent* (see Table I).

Equations (21) and (24) together with the definitions (10) and (23) form the content of the usual Gibbs-Helmholtz relations and provide grounds to call $\Phi_{eq}$ and $\Lambda_{eq}$ the *characteristic state* fns. in the corresponding variables (for closed and open system, respectively). Note that from the strict convexity of $\Lambda_{eq}$, Eq.(22), follows *positive semidefiniteness* (see Appendix B) of the second-derivatives matrix $\partial^2 \Lambda_{eq}[\alpha_F]/(\partial \alpha_i \partial \alpha_j)$. Since it is identical with the first-order covariance matrix, $\langle\langle \hat{O}_i, \hat{O}_j \rangle\rangle = (\partial^2 \Lambda_{eq}[\alpha_F, \hat{O}_F]/(\partial \alpha_i \partial \alpha_j))$ [differentiate both sides of Eq.(19) w.r.t. $\alpha_j$ and use the definition (3)], our result is in agreement with the well known positive semi-definiteness property of the covariance matrix. It should be noted that the positive definiteness of the matrix $\Lambda_{FF} = \{\partial^2 \Lambda_{eq}[\alpha_F]/(\partial\alpha_i \partial\alpha_j)\}$ implies the existence and positive definiteness of the matrix reciprocal to it, see Appendix B. Similarly, from the strict concavity of the action function $\Phi_{eq}[o_F]$ follows *negative semidefiniteness* of its second-derivatives matrix $\Phi_{FF} = \{\partial^2 \Phi_{eq}[o_F]/(\partial o_i \partial o_j)\}$. We obtain here explicitly this property of $\Phi_{eq}$ from the property of $\Lambda_{eq}$. Let us differentiate Eq.(24) w.r.t. $o_F$ to obtain

$$\Phi_{FF} = \left(\frac{\partial \alpha_F^{eq}}{\partial o_F}\right). \tag{32}$$

But $\alpha_F^{eq}[o_F]$ is the solution of Eq.(20). When differentiated w.r.t. $o_F$, the indicated equation leads to

$$\Lambda_{FF} \frac{\partial \alpha_F^{eq}}{\partial o_F} = -1_{FF}. \tag{33}$$

(here $1_{FF}$ denotes the unit matrix $\delta_{ij}$). This equation allows the interpretation of the matrix $-(\partial\alpha_F^{eq}/\partial o_F)$ as a reciprocal to the matrix $\Lambda_{FF}$ (provided it is a nonsingular matrix). Therefore, Eq.(32) can be rewritten as

$$\Phi_{FF} = -(\Lambda_{FF})^{-1} \tag{34}$$

## III. MASSIEU FUNCTIONS — PARTIAL LEGENDRE TRANSFORMS OF STATE FUNCTIONS.

Let us consider now the equilibrium state of a generic $(M,m)$ system with the (fixed number) $M$ degrees of freedom (i.e., $M$ generating operators $\{\hat{O}_j\}_{j=1}^M$) for which only $(M-m)$ expectation values have been imposed as constraints (with $0 \leq m \leq M$). The components of the vector $\hat{O}_F = (\hat{O}_1\ \hat{O}_2\ ...\ \hat{O}_M)^T$ are reordered to have first $m$ operators conjugate to the independent external sources $\{\alpha_j\}_{j=1}^m$, followed by $(M-m)$ operators with their expectation values $\{o_j\}_{j=m+1}^M$ imposed as the independent variables. The considered equilibrium state of the $(M,m)$ system (for a given order of components of the observable vector $\hat{O}_F$) will be named also the $(M,m)$ *thermodynamic ensemble* of the system $\{\hat{O}_j\}_{j=1}^M$. This name is a generalization of such traditional names as the grand canonical ensemble, canonical ensemble, etc. In the exemplary case of $M=2$, we take $m=2$. The independent variables are two external sources corresponding to observables $\hat{H}$ and $\hat{N}$: $\alpha_1 = \beta$ (the reciprocal temperature) and $\alpha_2$.

It will be convenient to introduce the sets of indices: $\mathsf{L} = \{1, 2, ..., m\}$, $\mathsf{U} = \{m+1, m+2, ..., M\}$, $\mathsf{F} = \{1, 2, ..., M\} = \mathsf{L} \cup \mathsf{U}$ of the lower ($\mathsf{L}$), upper ($\mathsf{U}$) and full ($\mathsf{F}$) range, respectively. Next, notations for the column vectors of variables (functions, operators) $x_\mathsf{L}$, $x_\mathsf{U}$, $x_\mathsf{F}$ are defined by: $x_\mathsf{L}^T = (x_1\ x_2\ ...\ x_m)^T$, $x_\mathsf{U}^T = (x_{m+1}\ x_{m+2}\ ...\ x_M)^T$, $x_\mathsf{F}^T = (x_\mathsf{L}^T\ \vdots\ x_\mathsf{U}^T) = (x_1 x_2 ...\ x_M)^T$. However, for the typographic reasons, the column vector $x_\mathsf{F}$ composed of subvectors $x_\mathsf{L}$ and $x_\mathsf{U}$ will be written alternatively as $x_\mathsf{F} = (x_\mathsf{L}, x_\mathsf{U})$ besides $x_\mathsf{F} = (x_\mathsf{L}^T\ \vdots\ x_\mathsf{U}^T)^T$. A notation for the matrix calculus used in this chapter is shortly recalled in Appendix A. In the exemplary case: $\mathsf{L} = \mathsf{F} = \{1, 2\}$, $\mathsf{U} = \varnothing$. So $x_\mathsf{L}$ is a two-component vector, $x_\mathsf{U}$ is empty.

The independent variables $(\alpha_\mathsf{L}, o_\mathsf{U})$ of the $(M,m)$ thermodynamic system can be considered as obtained via trivial mapping of the vector $\alpha_\mathsf{F} = (\alpha_\mathsf{L}, \alpha_\mathsf{U}) \in \mathcal{A}$, namely

$$(\alpha_\mathsf{L}, \alpha_\mathsf{U}) \mapsto (\alpha_\mathsf{L}, o_\mathsf{U}) = (\alpha_\mathsf{L}, o_\mathsf{U}^{eq}[\alpha_\mathsf{F}]) \equiv \mathcal{T}_{M,M}^{M,m}[\alpha_\mathsf{F}], \tag{35}$$

where, as it follows from the generating property of $\Lambda_{eq}$, Eqs.(20),(21),

$$\Lambda_{\mathrm{U}}[\alpha_{\mathrm{L}},\alpha_{\mathrm{U}}] = -o_{\mathrm{U}} = -o_{\mathrm{U}}^{\mathrm{eq}}[\alpha_{\mathrm{L}},\alpha_{\mathrm{U}}] \tag{36}$$

(see Appendix D for the description of $\mathcal{T}_y^x$).

The domain $\mathcal{D}^{M,m}$ of acceptable points $(\alpha_{\mathrm{L}}, o_{\mathrm{U}})$ is defined by the mapping (35) as

$$\mathcal{D}^{M,m} = \mathcal{T}_{M,M}^{M,m}[\mathcal{A}] \equiv \mathcal{T}_{M,M}^{M,m}[\mathcal{D}^{M,M}] \tag{37}$$

(see Table II).

The convenient thermodynamic fn. for the $(M,m)$ system is the Legendre transform $\Theta^{M,m}[\alpha_{\mathrm{L}}, o_{\mathrm{U}}]$ of $\Lambda_{\mathrm{eq}}[\alpha_{\mathrm{L}}, \alpha_{\mathrm{U}}]$ connected with Eq.(36) as the generating equation (see Appendix C)

$$\Theta^{M,m}[\alpha_{\mathrm{L}}, o_{\mathrm{U}}^{\mathrm{eq}}[\alpha_{\mathrm{L}}, \alpha_{\mathrm{U}}]] = \Lambda_{\mathrm{eq}}[\alpha_{\mathrm{L}}, \alpha_{\mathrm{U}}] - \left(-o_{\mathrm{U}}^{\mathrm{eq}}[\alpha_{\mathrm{L}}, \alpha_{\mathrm{U}}]\right)^{\mathrm{T}} \alpha_{\mathrm{U}}. \tag{38}$$

This function generates the reverse (to $\mathcal{T}_{M,M}^{M,m}$) mapping

$$(\alpha_{\mathrm{L}}, o_{\mathrm{U}}) \mapsto (\alpha_{\mathrm{L}}, \alpha_{\mathrm{U}}) = (\alpha_{\mathrm{L}}, \alpha_{\mathrm{U}}^{M,m}[\alpha_{\mathrm{L}}, o_{\mathrm{U}}]) \equiv \mathcal{T}_{M,m}^{M,M}[\alpha_{\mathrm{L}}, o_{\mathrm{U}}] \tag{39}$$

because it is connected with the first-derivatives vector, Eq.(C7),

$$\Theta_{\mathrm{U}}^{M,m}[\alpha_{\mathrm{L}}, o_{\mathrm{U}}] = \alpha_{\mathrm{U}}^{M,m}[\alpha_{\mathrm{L}}, o_{\mathrm{U}}]. \tag{40}$$

Therefore $\Lambda_{\mathrm{eq}}$ can be obtained as the Legendre transform of $\Theta^{M,m}$

$$\Lambda_{\mathrm{eq}}[\alpha_{\mathrm{L}}, \alpha_{\mathrm{U}}^{M,m}[\alpha_{\mathrm{L}}, o_{\mathrm{U}}]] = \Theta^{M,m}[\alpha_{\mathrm{L}}, o_{\mathrm{U}}] - \left(\alpha_{\mathrm{U}}^{M,m}[\alpha_{\mathrm{L}}, o_{\mathrm{U}}]\right)^{\mathrm{T}} o_{\mathrm{U}} \tag{41}$$

This means that $\Theta^{M,m}$ and $\Lambda_{\mathrm{eq}}$ are *Legendre-transform equivalent*. Interpretation of $\Theta^{2,2}[\alpha_1, \alpha_2]$ for the exemplary case in postponed to the next Section. The relations (38) and (41) expressed in terms of the transformed variables are

$$\Theta^{M,m}[\alpha_{\mathrm{L}}, o_{\mathrm{U}}] = \Lambda_{\mathrm{eq}}[\alpha_{\mathrm{L}}, \alpha_{\mathrm{U}}^{M,m}[\alpha_{\mathrm{L}}, o_{\mathrm{U}}]] + o_{\mathrm{U}}^{\mathrm{T}} \alpha_{\mathrm{U}}^{M,m}[\alpha_{\mathrm{L}}, o_{\mathrm{U}}], \tag{42}$$

$$\Lambda_{\mathrm{eq}}[\alpha_{\mathrm{L}}, \alpha_{\mathrm{U}}] = \Theta^{M,m}[\alpha_{\mathrm{L}}, o_{\mathrm{U}}^{\mathrm{eq}}[\alpha_{\mathrm{L}}, \alpha_{\mathrm{U}}]] - \alpha_{\mathrm{U}}^{\mathrm{T}} o_{\mathrm{U}}^{\mathrm{eq}}[\alpha_{\mathrm{L}}, \alpha_{\mathrm{U}}]. \tag{43}$$

The first derivatives of these two functions w.r.t. $\alpha_L$ are equal, Eq.(C4), and they can be written in two versions, Eqs.(C10), (C11),

$$\Theta_L^{M,m}\left[\alpha_L, o_U\right] = \Lambda_L\left[\alpha_L, \alpha_U^{M,m}\left[\alpha_L, o_U\right]\right] = -o_L^{eq}\left[\alpha_L, \alpha_U^{M,m}\left[\alpha_L, o_U\right]\right] \quad \forall \left(\alpha_L, o_U\right) \in \mathcal{D}^{M,m}, \quad (44)$$

$$\Theta_L^{M,m}\left[\alpha_L, o_U^{eq}\left[\alpha_L, \alpha_U\right]\right] = \Lambda_L\left[\alpha_L, \alpha_U\right] = -o_L^{eq}\left[\alpha_L, \alpha_U\right] \qquad \forall \alpha_F \in \mathcal{A}. \quad (45)$$

The independent variables of the $(M, m)$ system can be considered alternatively as obtained via trivial mapping of the vector $o_F = \left(o_L, o_U\right) \in \mathcal{O} \equiv \mathcal{D}^{M,0}$

$$\left(o_L, o_U\right) \mapsto \left(\alpha_L, o_U\right) = \left(\alpha_L^{eq}\left[o_L, o_U\right], o_U\right) \equiv \mathcal{T}_{M,0}^{M,m}\left[o_L, o_U\right], \quad (46)$$

where the generating property of $\Phi_{eq}$, Eq.(24), gives

$$\Phi_L\left[o_L, o_U\right] = \alpha_L = \alpha_L^{eq}\left[o_L, o_U\right]. \quad (47)$$

The Legendre transform of $\Phi_{eq}\left[o_L, o_U\right]$ connected with the generating Eq.(47) is

$$^\Phi\Theta^{M,m}\left[\alpha_L^{eq}\left[o_L, o_U\right], o_U\right] = \Phi_{eq}\left[o_L, o_U\right] - o_L^T \alpha_L^{eq}\left[o_L, o_U\right]. \quad (48)$$

This fn. generates the reverse (to $\mathcal{T}_{M,0}^{M,m}$) mapping

$$\left(\alpha_L, o_U\right) \mapsto \left(o_L, o_U\right) = \left(o_L^{M,m}\left[\alpha_L, o_U\right], o_U\right) \equiv \mathcal{T}_{M,m}^{M,0}\left[\alpha_L, o_U\right], \quad (49)$$

because of its connection with the first derivatives vector, Eq.(C7),

$$^\Phi\Theta_L^{M,m}\left[\alpha_L, o_U\right] = -o_L^{M,m}\left[\alpha_L, o_U\right]. \quad (50)$$

Therefore $\Phi_{eq}$ can be obtained as the Legendre transform of $^\Phi\Theta^{M,m}$

$$\Phi_{eq}\left[o_L^{M,m}\left[\alpha_L, o_U\right], o_U\right] = {}^\Phi\Theta^{M,m}\left[\alpha_L, o_U\right] + \alpha_L^T o_L^{M,m}\left[\alpha_L, o_U\right]. \quad (51)$$

This means that $^\Phi\Theta^{M,m}$ and $\Phi_{eq}$ are Legendre-transform equivalent. The relations (48) and (51) in terms of the transformed variables are

$$^\Phi\Theta^{M,m}\left[\alpha_L, o_U\right] = \Phi_{eq}\left[o_L^{M,m}\left[\alpha_L, o_U\right], o_U\right] - \alpha_L^T o_L^{M,m}\left[\alpha_L, o_U\right], \quad (52)$$

$$\Phi_{eq}\left[o_L,o_U\right] = {}^{\Phi}\Theta^{M,m}\left[\alpha_L^{eq}\left[o_L,o_U\right],o_U\right] + o_L^T\alpha_L^{eq}\left[o_L,o_U\right]. \tag{53}$$

The first derivatives of these two functions w.r.t. $o_U$ are equal, Eq.(C.4), and they can be written in two versions

$$^{\Phi}\Theta_U^{M,m}\left[\alpha_L,o_U\right] = \Phi_U\left[o_L^{M,m}\left[\alpha_L,o_U\right],o_U\right] = \alpha_U^{eq}\left[o_L^{M,m}\left[\alpha_L,o_U\right],o_U\right] \quad \forall\left(\alpha_L,o_U\right)\in\mathcal{D}^{M,m}, \tag{54}$$

$$^{\Phi}\Theta_U^{M,m}\left[\alpha_L^{eq}\left[o_L,o_U\right],o_U\right] = \Phi_U\left[o_L,o_U\right] = \alpha_U^{eq}\left[o_L,o_U\right] \quad \forall o_F \in \mathcal{O}. \tag{55}$$

Since the $(M,m)$ system is unique, its independent variables obtained directly from the fully open system variables, $\left(\alpha_L,o_U\right) = \mathcal{T}_{M,M}^{M,m}\left[\alpha_L,\alpha_U\right]$, Eq.(35), should be the same as obtained via the fully closed system variables, $\left(\alpha_L,o_U\right) = \mathcal{T}_{M,0}^{M,m}\left[\mathcal{T}_{M,M}^{M,0}\left[\alpha_L,\alpha_U\right]\right] = \mathcal{T}_{M,0}^{M,m}\left[o_F^{eq}\left[\alpha_L,\alpha_U\right]\right]$, Eqs.(21), (46), Table II, so equality

$$\left(\alpha_L,o_U^{eq}\left[\alpha_L,\alpha_U\right]\right) = \left(\alpha_L^{eq}\left[o_F^{eq}\left[\alpha_L,\alpha_U\right]\right],o_U^{eq}\left[\alpha_L,\alpha_U\right]\right) \quad \forall\left(\alpha_L,\alpha_U\right)\in\mathcal{A} \tag{56}$$

should hold. Really, it is true due to Eq.(27). After applying this identity (27) to Eq.(48) in which $o_F$ is substituted by $o_F^{eq}\left[\alpha_F\right]$ we find

$$^{\Phi}\Theta^{M,m}\left[\alpha_L,o_U^{eq}\left[\alpha_L,\alpha_U\right]\right] = \Phi_{eq}\left[o_F^{eq}\left[\alpha_F\right]\right] - \alpha_L^T o_L^{eq}\left[\alpha_F\right]. \tag{57}$$

Next, with $\Phi_{eq}$ obtained from Eq.(31) as

$$\Phi_{eq}\left[o_F^{eq}\left[\alpha_F\right]\right] = \Lambda_{eq}\left[\alpha_F\right] + \alpha_L^T o_L^{eq}\left[\alpha_F\right] + \alpha_U^T o_U^{eq}\left[\alpha_F\right] \tag{58}$$

and substituted to Eq.(57), we see that $^{\Phi}\Theta^{M,m}$ given in Eq.(57) is identical to $\Theta^{M,m}$ given in Eq.(38). Thus, all relations obtained for $^{\Phi}\Theta^{M,m}$, Eqs.(50)–(55), hold for $\Theta^{M,m}$. Two Eqs. (50) and (40) can be combined into one equation

$$\Theta_F^{M,m}\left[\alpha_L,o_U\right] = \left(-o_L^{M,m}\left[\alpha_L,o_U\right],\alpha_U^{M,m}\left[\alpha_L,o_U\right]\right). \tag{59}$$

From the mappings (39) and (49) follow two equivalent forms of the eq-DM of the $(M,m)$ system

$$\hat{\Gamma}_{eq}^{M,m}[\alpha_L, o_U] = \hat{\Gamma}_{eq}[\alpha_L, \alpha_U^{M,m}[\alpha_L, o_U]] = \hat{\Gamma}_{eq}^{av}[o_L^{M,m}[\alpha_L, o_U], o_U]. \qquad (60)$$

The functions $\Phi_{eq}[o_L, o_U]$ (for $0 \leq m \leq M$) will be named the state functions in the entropy representation.

In summary, the completely open system is characterized by $\Theta^{M,M}[\alpha_F] = \Lambda_{eq}[\alpha_F]$ (U is the empty set), the independent variables are all external sources. The fully isolated system is characterized by $\Theta^{M,0}[o_F] = \Phi_{eq}[o_F]$ (L is the empty set), the independent variables are all expectation values. The partially open/closed system, $0 < m < M$, is characterized by the fn. $\Theta^{M,m}[\alpha_L, o_U]$ — it will be named the $(M, m)$ *Massieu fn.* (see Table I). It is easy to verify that the conditions operator (a generalization of Eq.(5))

$$\hat{O}^{M,m}[\alpha_L, o_U] = -(\alpha_L, \alpha_U^{M,m}[\alpha_L, o_U])^T (\hat{O}_L, (\hat{O}_U - o_U)) \qquad (61)$$

generates directly the $(M, m)$ Massieu fn.

$$\Theta^{M,m} = \ln\left(\text{Tr}\exp(\hat{O}^{M,m})\right) \qquad (62)$$

[compare Eqs.(11) and (9)] and, when inserted to Eq.(8) it also leads to the same eq-DM as in Eq.(60),

$$\hat{\Gamma}_{eq}^{M,m} = \hat{\Gamma}_{eq}[\hat{O}^{M,m}] = \exp(\hat{O}^{M,m})/\text{Tr}\exp(\hat{O}^{M,m}), \qquad (63)$$

due to cancellations of the contributions $o_U^T \alpha_U^{M,m}$ in the numerator and denominator in Eq.(8). According to Eq.(28), the effective action function $\Phi_{eq}[o_L^{M,m}, o_U]$ in Eq.(52) can be replaced by the entropy $S^{BGS}[\hat{\Gamma}_{eq}^{M,m}]$.

The matrix $\Theta_{FF}^{M,m}$ of second derivatives of $\Theta^{M,m}[\alpha_L, o_U]$ with respect to its arguments vector is obtained in Appendix A in two equivalent forms in terms of submatrices of either $\Lambda_{FF} = \partial^2 \Lambda_{eq}[\alpha_F]/(\partial \alpha_F)^2$ or $\Phi_{FF} = \partial^2 \Phi_{eq}[o_F]/(\partial o_F)^2$, namely

$$\Theta_{FF}^{M,m} \equiv \frac{\partial^2 \Theta^{M,m}[\alpha_L, o_U]}{(\partial(\alpha_L, o_U))^2} = \begin{pmatrix} \Lambda_{LL} - \Lambda_{LU}\Lambda_{UU}^{-1}\Lambda_{UL} & -\Lambda_{LU}\Lambda_{UU}^{-1} \\ -\Lambda_{UU}^{-1}\Lambda_{UL} & -\Lambda_{UU}^{-1} \end{pmatrix}$$
$$= \begin{pmatrix} -\Phi_{LL}^{-1} & \Phi_{LL}^{-1}\Phi_{LU} \\ \Phi_{UL}\Phi_{LL}^{-1} & \Phi_{UU} - \Phi_{UL}\Phi_{LL}^{-1}\Phi_{LU} \end{pmatrix}. \quad (64)$$

Note that $\Theta_{FF}^{M,m}$ is symmetric, as it should be, e.g. $(\Phi_{UL}\Phi_{LL}^{-1})^T = (\Phi_{LL}^{-1})^T \Phi_{UL}^T = \Phi_{LL}^{-1}\Phi_{LU}$ — the last step due to the symmetry of $\Phi_{FF}$.

Remembering that $\partial^2 \Lambda_{eq}[\alpha_F]/(\partial \alpha_i \partial \alpha_j) = \langle\langle \hat{O}_i, \hat{O}_j \rangle\rangle$, i.e., $\Lambda_{AB} = \langle\langle \hat{O}_A, \hat{O}_B^T \rangle\rangle$ for the blocks, Eq.(64) can be rewritten in terms of the first-order covariance matrix as

$$\Theta_{FF}^{M,m} = \begin{pmatrix} \langle\langle \hat{O}_L, \hat{O}_L^T \rangle\rangle - \langle\langle \hat{O}_L, \hat{O}_U^T \rangle\rangle \langle\langle \hat{O}_U, \hat{O}_U^T \rangle\rangle^{-1} \langle\langle \hat{O}_U, \hat{O}_L^T \rangle\rangle & -\langle\langle \hat{O}_L, \hat{O}_U^T \rangle\rangle \langle\langle \hat{O}_U, \hat{O}_U^T \rangle\rangle^{-1} \\ -\langle\langle \hat{O}_U, \hat{O}_U^T \rangle\rangle^{-1} \langle\langle \hat{O}_U, \hat{O}_L^T \rangle\rangle & -\langle\langle \hat{O}_U, \hat{O}_U^T \rangle\rangle^{-1} \end{pmatrix}. \quad (65)$$

The diagonal blocks of $\Theta_{FF}^{M,m}$ have particular definiteness: $\Theta_{LL}^{M,m}$ is positive definite, $\Theta_{UU}^{M,m}$ is negative definite. To find these properties, we recall from Appendix B that from negative definiteness of $\Phi_{FF}$ follows the same property of $\Phi_{LL}$ and therefore of $\Phi_{LL}^{-1}$, finally $\Theta_{LL}^{M,m} = -\Phi_{LL}^{-1}$ is positive definite. Similarly, from positive definiteness of $\Lambda_{FF}$ follows negative definiteness of $\Theta_{UU}^{M,m} = -\Lambda_{UU}^{-1}$. (While, in general, the matrices $\Phi_{FF}$ and $\Lambda_{FF}$ have only particular *semidefinitness*, the last two conclusions concern their argument points at which matrices $\Phi_{LL}$ and $\Lambda_{UU}$ are not singular, i.e., have particular *definiteness*. Next, the diagonal blocks of $\Theta_{FF}^{M,m}$ have particular definiteness rather than semidefiniteness, because *infinite* eigenvalues of $\Phi_{LL}$ and $\Lambda_{UU}$ must be excluded as unphysical). The established properties of $\Theta^{M,m}_{LL}$ and $\Theta^{M,m}_{UU}$ can be summed up as *strict convexity* of $\Theta^{M,m}[\alpha_L, o_U]$ w.r.t $\alpha_L$ and *strict concavity* w.r.t. $o_U$.

As a concave function of $o_U$ (at fixed $\alpha_L$), $\Theta^{M,m}[\alpha_L, o_U]$ can reach its maximum at some unique $o_U = o_U^0[\alpha_L]$. Obviously, at this point, the first derivatives vanish, $\Theta^{M,m}_U[\alpha_L, o_U^0[\alpha_L]] = 0_U$ and, therefore, $\alpha^{M,m}_U[\alpha_L, o_U^0[\alpha_L]] = 0_U$ follows from Eq.(59) (all components of the vector $0_U$ are zero). After inserting this result $\alpha^{M,m}_U = 0_U$ into Eq.(61) we see that the upper-range vector operator $\hat{O}_U$ of the total generating vector operator $\hat{O}_F = (\hat{O}_L, \hat{O}_U)$ is, in fact, not involved in the general conditions operators $\hat{O}^{M,m}$, therefore the number of degrees of freedom in the considered state of the thermodynamic systems is reduced from $M$ to $m$ (as specified by the components of $\hat{O}_L$). Of course, one can consider also a state of this system having another number $k$ of its first derivatives w.r.t. $o_j$ equal zero, $1 \le k \le (M-m)$. Then, at equilibrium, this system is equivalent system with the number of degrees of freedom reduced to $(M-k)$.

## IV. GIBBS-HELMHOLTZ FUNCTIONS — THE MASSIEU-PLANCK TRANSFORMS OF MASSIEU FUNCTIONS.

To be closer to traditional thermodynamic functions used by chemical and physical communities, which prefer to use the criteria based on Gibbs and Helmholtz potential, we define now the $(M,m)$ *Gibbs-Helmholtz fn.* $\Upsilon^{M,m}$ as the *Massieu-Planck transform* [65] (for $1 \le m \le M$) of the $(M,m)$ Massieu fn. $\Theta^{M,m}$ given by

$$\Upsilon^{M,m}[\beta, \gamma_{L'}, o_U] = -\beta^{-1} \Theta^{M,m}[\beta, -\beta\gamma_{L'}, o_U], \tag{66}$$

i.e., by introducing a prefactor $(-\beta^{-1})$ at $\Theta^{M,m}$ and inserting to it the specific external sources

$$\hat{\alpha}_{\mathsf{L}} \equiv \hat{\alpha}_{\mathsf{L}}\left[\beta, \gamma_{\mathsf{L}'}\right] = \beta\left(1, -\gamma_{\mathsf{L}'}\right), \tag{67}$$

(here $\mathsf{L}' = \{2,...,m\}$, similarly $\mathsf{F}' = \{2,...,M\}$). Using Eq.(42) and (52), this $\Upsilon^{M,m}$ can be also rewritten as

$$\begin{aligned}
\Upsilon^{M,m}\left[\beta, \gamma_{\mathsf{L}'}, o_{\mathsf{U}}\right] &= -\beta^{-1}\Lambda_{eq}\left[\hat{\alpha}_{\mathsf{L}}, \alpha_{\mathsf{U}}^{M,m}\right] + o_{\mathsf{U}}^{\mathsf{T}}\gamma_{\mathsf{U}}^{M,m} \\
&= -\beta^{-1}\Phi_{eq}\left[o_{\mathsf{L}}^{M,m}, o_{\mathsf{U}}\right] + o_1^{M,m} - \gamma_{\mathsf{L}'}^{\mathsf{T}} o_{\mathsf{L}'}^{M,m} \\
&= -\beta^{-1}S^{\mathrm{BGS}}\left[\beta, \gamma_{\mathsf{L}'}, o_{\mathsf{U}}\right] + o_1^{M,m} - \gamma_{\mathsf{L}'}^{\mathsf{T}} o_{\mathsf{L}'}^{M,m},
\end{aligned} \tag{68}$$

where we defined $S^{\mathrm{BGS}}\left[\beta, \gamma_{\mathsf{L}'}, o_{\mathsf{U}}\right] = S^{\mathrm{BGS}}\left[\hat{\Gamma}\left[\hat{O}^{M,m}\left[\beta, -\beta\gamma_{\mathsf{L}'}, o_{\mathsf{U}}\right]\right]\right]$, see Eq.(61), and

$$\gamma_{\mathsf{U}}^{M,m} \equiv \gamma_{\mathsf{U}}^{M,m}\left[\beta, \gamma_{\mathsf{L}'}, o_{\mathsf{U}}\right] = -\beta^{-1}\alpha_{\mathsf{U}}^{M,m}\left[\hat{\alpha}_{\mathsf{L}}, o_{\mathsf{U}}\right]. \tag{69}$$

In thermodynamic applications, $\beta$ means the reciprocal temperature (in an energy unit), $\beta > 0$, being the external source conjugate to the expectation value $o_1$ of some energy operator $\hat{O}_1$ (like the total energy, the internal energy, etc.). Therefore $\Upsilon^{M,m}$ is a quantity with the dimension of the energy. For the exemplary case we take $\gamma_2 = \mu$ and then find $\Upsilon^{2,2}[\beta, \mu] = \Omega^0[\beta, \mu]$ – the (2,2) Gibbs-Helmholtz fn. coincides with the grand potential of the grand-canonical ensemble. The independent parameter $\gamma_2 = \mu$ is known as the chemical potential of the system. The explicit form of the grand potential is $\Omega^0[\beta, \mu] = E[\beta, \mu] - \mu N[\beta, \mu] - \beta^{-1}S^{\mathrm{BGS}}[\beta, \mu]$. Finally, the (2,2) Massieu fn. $\Theta^{2,2}[\alpha_1 = \beta, \alpha_2 = -\beta\mu]$ of previous section is the reverese Massieu-Planck transform of $\Omega^0[\beta, \mu]$. In general, for the distinctive external source $\beta$ one can choose any component $\alpha_i$ of the vector $\alpha_{\mathsf{L}}$ and then reorder the components appropriately to obtain again $\Upsilon^{M,m}$ in the form (66), having the same dimension as that of the chosen $\hat{O}_i$; acceptable sign of $\beta$ depends on the nature of $\hat{O}_i$. Note that to each ($M$, $m$) Massieu fn. $\Theta^{M,m}$ with $1 \leq m \leq M$ there corresponds a set of $m$ Gibbs-Helmholtz fns., $\Upsilon^{M,m}$ (any $\alpha_i$ from $\alpha_{\mathsf{L}}$ can be chosen to be $\beta$). Similarly to the property exhibited by $\Theta^{M,m}$, the $(M, m)$ Gibbs-Helmholtz fn. is a convenient generating function for the considered system. As shown in Appendix A, its first derivative is

$$\Upsilon_F^{M,m}\left[\beta,\gamma_{L'},o_U\right] = \left(\frac{\partial \Upsilon^{M,m}\left[\beta,\gamma_{L'},o_U\right]}{\partial(\beta,\gamma_{L'},o_U)}\right)^T = \left(\beta^{-2}S^{BGS}\left[\beta,\gamma_{L'},o_U\right], -o_{L'}^{M,m}, \gamma_U^{M,m}\right), \quad (70)$$

where $\hat{\alpha}_L = (\beta, -\beta\gamma_{L'})$ should be inserted for the argument $\alpha_L$ of $o_{L'}^{M,m}$ and of $\gamma_U^{M,m}$ in the above result. The matrix of the second derivatives, obtained in Appendix A, is

$$\Upsilon_{FF}^{M,m} \equiv \frac{\partial^2 \Upsilon^{M,m}\left[\beta,\gamma_{L'},o_U\right]}{\left(\partial(\beta,\gamma_{L'},o_U)\right)^2} = -\beta^{-1}\left(\begin{pmatrix} 2\beta^{-2}S^{BGS} & 0_{L'}^T & \left(\gamma_U^{M,m}\right)^T \\ 0_{L'} & & 0_{F'F'} \\ \gamma_U^{M,m} & & \end{pmatrix} + y_{F,F}^T \Theta_{FF}^{M,m} y_{F,F}\right), \quad (71)$$

where the Jacobian matrix $y_{F,F}\left[\beta,\gamma_{L'},o_U\right]$ is given in Eq.(A23). It should be noted that the above second-derivatives matrix $\Upsilon_{FF}^{M,m}$ of the fn. $\Upsilon^{M,m}$ is written in terms of the second-derivatives matrix $\Theta_{FF}^{M,m}$ of the fn. $\Theta^{M,m}$, being therefore much more complicated expression than $\Theta_{FF}^{M,m}$ itself. For the exemplary case of $L = F = \{1,2\}$, we have the first-derivative vector

$$\Upsilon_F^{2,2}[\beta,\mu] = \left(\beta^{-2}S^{BGS}[\beta,\mu], -N[\beta,\mu]\right)^T \quad (72)$$

and the second-derivative matrix

$$\Upsilon_{FF}^{2,2}[\beta,\mu] = -\beta^{-1}\left(2\beta^{-2}S^{BGS}\begin{pmatrix} 1 & 0 \\ 0 & 0 \end{pmatrix} + y_{F,F}^T \Theta_{FF}^{M,m} y_{F,F}\right), \quad (73)$$

where

$$y_{F,F}[\beta,\mu] = \begin{pmatrix} 1 & 0 \\ -\mu & -\beta \end{pmatrix}, \quad (74)$$

$$\Theta_{FF}^{2,2}[\alpha_1 = \beta, \alpha_2 = -\beta\mu] = \Lambda_{FF} = \begin{pmatrix} \langle\langle \hat{H}, \hat{H} \rangle\rangle & \langle\langle \hat{H}, \hat{N} \rangle\rangle \\ \langle\langle \hat{N}, \hat{H} \rangle\rangle & \langle\langle \hat{N}, \hat{N} \rangle\rangle \end{pmatrix}. \quad (75)$$

It should be noted that $\Theta_{FF}^{2,2}$ is the second-derivatives matrix of the (2,2) Massieu fn. of the previous Section. Its firs-derivative vector is $\Theta_F^{2,2}[\alpha_1,\alpha_2] = -(E[\alpha_1,\alpha_2], N[\alpha_1,\alpha_2])^T$.

To analyze the stability criteria of the Gibbs-Helmholtz fn. $\Upsilon^{M,m}$ we evaluate two diagonal blocks of its second-derivatives matrix, Eq.(71):

$$-\beta \Upsilon_{LL}^{M,m} = 2\beta^{-2} S^{BGS} \left( \begin{array}{c|c} 1 & 0_{L'}^T \\ \hline 0_{L'} & 0_{L'L'} \end{array} \right) + y_{L,L}^T \Theta_{LL}^{M,m} y_{L,L}, \tag{76}$$

$$-\beta \Upsilon_{UU}^{M,m} = \Theta_{UU}^{M,m}, \tag{77}$$

and, for arbitrary nonzero vector $v_L$, the scalar $-\beta v_L^T \Upsilon_{LL}^{M,m} v_L = 2\beta^{-2} S^{BGS} v_1^2 + (y_{L,L} v_L)^T \Theta_{LL}^{M,m} (y_{L,L} v_L)$ based on Eq.(76). Its first term is nonnegative due to $S^{BGS} \geq 0$. Since $\det(y_{L,L}) = (-z_1)^{m-1} = (-\beta)^{m-1} \neq 0$, the vector $y_{L,L} v_L$ is nonzero too. Due to this fact and the positive definiteness of $\Theta_{LL}^{M,m}$, the second term of the scalar is positive [see Eq.(B1)]. As the considered scalar is positive, $\Upsilon_{LL}^{M,m}$ in Eq.(76) is negative definite when $\beta > 0$, according to Appendix B. Due to the negative definiteness of $\Theta_{UU}^{M,m}$, the block $\Upsilon_{UU}^{M,m}$ of Eq.(77) is positive definite when $\beta > 0$. Therefore, $\Upsilon^{M,m}[\beta, \gamma_{L'}, o_U]$ is a *strictly concave* fn. of $(\beta, \gamma_{L'})$ and a *strictly convex* fn. of $o_U$. When $\beta < 0$ (possible when the external source $\beta$ is not the reciprocal temperature), the properties of the definiteness and curvature are to be reversed. Since the first variable, $\beta$, is treated differently than $\gamma_{L'}$ in the definition of $\Upsilon^{M,m}[\beta, \gamma_{L'}, o_U]$, Eq.(66), we evaluate also the corresponding blocks of $\Upsilon_{LL}^{M,m}$ to find

$$\Upsilon_{11}^{M,m} = -\beta^{-3} \left( 2 S^{BGS} + \widehat{\alpha}_L^T \Theta_{LL}^{M,m} \widehat{\alpha}_L \right), \tag{78}$$

$$\Upsilon_{L'L'}^{M,m} = -\beta^{+1} \Theta_{L'L'}^{M,m}. \tag{79}$$

Using Eq.(65), the blocks $\Upsilon_{LL}^{M,m}$ and $\Upsilon_{UU}^{M,m}$ can be easily rewritten in terms of the first-order covariance matrix.

The Massieu-Planck transformation of $\Theta^{M,m}$ into $\Upsilon^{M,m}$, Eq.(66), can be reversed uniquely as

$$\Theta^{M,m}\left[\beta,\alpha_{L'},o_U\right]=-\beta\Upsilon^{M,m}\left[\beta,-\beta^{-1}\alpha_{L'},o_U\right]. \tag{80}$$

Therefore the state fns. $\Upsilon^{M,m}$ and $\Theta^{M,m}$ are equivalent characteristics of the $(M,m)$ thermodynamic system for $1\leq m\leq M$ (see Table I).

Although the curvature properties of the $\Upsilon^{M,m}$ are the same as in the case of $\Theta^{M,m}$ (for $\beta<0$, and reversed for $\beta>0$) the functional dependence of $\Upsilon^{M,m}$ on $\beta$ is quite different than that of $\Theta^{M,m}$. Due to this, two descriptions of the $(M,m)$ system may reveal different aspects of its properties. In particular, the Massieu fn. $\Theta^{M,m}\left[\beta,-\beta\gamma_{L'},o_U\right]=\Theta^{M,m}\left[\beta v_L,o_U\right]$, where $v_L=(1,-\gamma_{L'})$ is fixed, attains its extremum at same point $\beta^0$ where its first derivative w.r.t. $\beta$

$$\left(\frac{\partial\Theta^{M,m}\left[\beta v_L,o_U\right]}{\partial\beta}\right)_{v_L,o_U}=v_L^T\Theta_L^{M,m}\left[\beta v_L,o_U\right]=-v_L^T o_L^{M,m}\left[\beta v_L,o_U\right] \tag{81}$$

vanishes. Thus $\beta=\beta^0\left[\gamma_{L'},o_U\right]$ is a solution of the equation

$$-o_1^{M,m}\left[\beta^0,-\beta^0\gamma_{L'},o_U\right]+\gamma_{L'}^T o_{L'}^{M,m}\left[\beta^0,-\beta^0\gamma_{L'},o_U\right]=0. \tag{82}$$

This extremum is, in fact, the minimum, because $\left(\partial^2\Theta^{M,m}\left[\beta v_L,o_U\right]/\partial\beta^2\right)_{v_L,o_U}=v_L^T\Theta_{LL}^{M,m}v_L>0$ due to positive definiteness of $\Theta_{LL}^{M,m}$.

On the other hand, the $(M,m)$ Gibbs-Helmholtz fn. $\Upsilon^{M,m}\left[\beta,\gamma_{L'},o_U\right]$ is a monotonically increasing fn. of $\beta$, because its derivative w.r.t $\beta$ (see Eq.(70))

$$\frac{\partial\Upsilon^{M,m}\left[\beta,\gamma_{L'},o_U\right]}{\partial\beta}=\frac{S^{BGS}\left[\hat{\Gamma}_{eq}^{M,m}\right]}{\beta^2} \tag{83}$$

is positive for $0 < \beta^2 < \infty$. If the zero value of the limit of this derivative, for $\beta^2 \to \infty$, exists, then $\Upsilon^{M,m}[\beta, \gamma_{L'}, o_U]$ attains a finite-value maximum for $\beta \to +\infty$ or minimum for $\beta \to -\infty$. The functions $\Upsilon^{M,m}[\beta, \gamma_{L'}, o_U]$ (for $1 \leq m \leq M$) will be named the state functions in the energy representation.

## V. CONCLUSIONS

The equilibrium state, which stems from the maximum entropy principle and is characterized by the basic Massieu fn. was defined and investigated. Concavity and convexity properties of the basic Massieu fn., its Legendre transforms and corresponding Massieu-Planck transforms for various ensembles were determined using the effective action formalism. Table I displays relations between various state functions defined and discussed in the paper. Since all the displayed functions are mutually connected with some direct and reverse transformation, they represent equivalent descriptors of the equilibrium state of the $\hat{O}_F = \{\hat{O}_1, \hat{O}_2, \ldots, \hat{O}_M\}$ thermodynamic system, for various possible sets of $M$ independent variables, chosen among elements of the external sources $\alpha_F$ and expectation values $o_F$. Since the arguments used here are general (the maximum entropy principle, the effective action formalism), the results obtained in the framework of the ensemble approach can provide a rigorous mathematical foundation for generalizations (to finite temperatures and to fractional number of electrons, spins, pairs) of various extension of DFT. In particular, the spin-density functional theory extension and the analysis of the zero-temperature limit of this extension will be the subject of the next two papers in the series.[74,75]

## ACKNOWLEDGMENT

The authors acknowledge support of Ministry of Science and Higher Education under Grant No. 1 T09A 025 30. Special thanks are due to Professor Paul W. Ayers for his valuable comments and beneficial remarks on the manuscript.

# APPENDIX A: VECTORS, BLOCK MATRICES, DERIVATIVES OF THERMODYNAMIC FUNCTIONS

A column matrix — a vector in $M$-dimensional space — is denoted by a variable with a subscript which indicates an ordered set of indices, e.g., $x_\mathsf{F}$. A row matrix is the transpose of a column matrix $x_\mathsf{F}^\mathrm{T} = (x_1 x_2 ... x_M)$. Here $\mathsf{F}$ means the full range of indices in the set, $\mathsf{F} = \{1, 2, ..., M\}$. It may be partitioned into subsets, e.g. the lower range $\mathsf{L} = \{1, 2, ..., m\}$ and the upper one $\mathsf{U} = \{m+1, ..., M\}$, $\mathsf{F} = \mathsf{L} \cup \mathsf{U}$. The corresponding block structure of a row matrix is denoted as $x_\mathsf{F}^\mathrm{T} = (x_\mathsf{L}^\mathrm{T} \mid x_\mathsf{U}^\mathrm{T})$. For typographic reasons, the corresponding column $x_\mathsf{F} = (x_\mathsf{L}^\mathrm{T} \mid x_\mathsf{U}^\mathrm{T})^\mathrm{T}$ may be alternatively denoted as $x_\mathsf{F} = (x_\mathsf{L}, x_\mathsf{U})$. A square or rectangular matrix is distinguished by two set subscripts, e.g. a square matrix $\Phi_\mathsf{FF}$ can be composed of its blocks: square ones $\Phi_\mathsf{LL}, \Phi_\mathsf{UU}$, and rectangular ones $\Phi_\mathsf{LU}$ and $\Phi_\mathsf{UL}$, namely

$$\Phi_\mathsf{FF} = \begin{pmatrix} \Phi_\mathsf{LL} & \Phi_\mathsf{LU} \\ \Phi_\mathsf{UL} & \Phi_\mathsf{UU} \end{pmatrix} = \{\Phi_{ij}\}_{i,j=1}^{M} \tag{A1}$$

and so on for higher-rank matrices like $\Psi_\mathsf{FFF}$ etc.

The gradient of a scalar fn. $f$ of a vector $x_\mathsf{F}$ is a row matrix

$$\frac{\partial f[x_\mathsf{F}]}{\partial x_\mathsf{F}} = \begin{pmatrix} \dfrac{\partial f}{\partial x_1} & \dfrac{\partial f}{\partial x_2} & \cdots & \dfrac{\partial f}{\partial x_M} \end{pmatrix} = (f_\mathsf{F}[x_\mathsf{F}])^\mathrm{T}, \qquad f_i = \frac{\partial f[x_\mathsf{F}]}{\partial x_i}. \tag{A2}$$

The matrix of second derivatives of this fn. is a symmetric square matrix, being the gradient of the first derivative vector

$$f_\mathsf{FF}[x_\mathsf{F}] = \frac{\partial^2 f[x_\mathsf{F}]}{(\partial x_\mathsf{F})^2} = \frac{\partial f_\mathsf{F}[x_\mathsf{F}]}{\partial x_\mathsf{F}} = \left\{\frac{\partial^2 f}{\partial x_j \partial x_i}\right\}_{i,j=1}^{M} = \{f_{ij}\}_{i,j=1}^{M}. \tag{A3}$$

Similarly, the Jacobian matrix, which describes locally the transformation $x_\mathsf{F} \to y_\mathsf{F}[x_\mathsf{F}]$ of vector variables, is given by

$$y_{\mathsf{F},\mathsf{F}}[x_{\mathsf{F}}] = \frac{\partial y_{\mathsf{F}}[x_{\mathsf{F}}]}{\partial x_{\mathsf{F}}} = \begin{pmatrix} \frac{\partial y_1}{\partial x_1} & \frac{\partial y_1}{\partial x_2} & \cdots \\ \frac{\partial y_2}{\partial x_1} & \frac{\partial y_2}{\partial x_2} & \cdots \\ \vdots & \vdots & \ddots \end{pmatrix}. \tag{A4}$$

Its typical application is the chain rule, e.g. finding the gradient of the fn. $f[x_{\mathsf{F}}] = g[y_{\mathsf{F}}[x_{\mathsf{F}}]]$, as

$$\frac{\partial f[x_{\mathsf{F}}]}{\partial x_{\mathsf{F}}} = \left(f_{\mathsf{F}}[x_{\mathsf{F}}]\right)^{\mathsf{T}} = \left(\frac{\partial g[y_{\mathsf{F}}]}{\partial y_{\mathsf{F}}}\right)\bigg|_{y_{\mathsf{F}} = y_{\mathsf{F}}[x_{\mathsf{F}}]} \frac{\partial y_{\mathsf{F}}[x_{\mathsf{F}}]}{\partial x_{\mathsf{F}}} = \left(g_{\mathsf{F}}[y_{\mathsf{F}}[x_{\mathsf{F}}]]\right)^{\mathsf{T}} y_{\mathsf{F},\mathsf{F}}[x_{\mathsf{F}}]. \tag{A5}$$

The gradient of a scalar product of the vectors $f_{\mathsf{F}}[x_{\mathsf{F}}]$ and $h_{\mathsf{F}}[x_{\mathsf{F}}]$ is

$$\frac{\partial}{\partial x_{\mathsf{F}}}\left(f_{\mathsf{F}}^{\mathsf{T}} h_{\mathsf{F}}\right) = \frac{\partial}{\partial x_{\mathsf{F}}}\left(h_{\mathsf{F}}^{\mathsf{T}} f_{\mathsf{F}}\right) = f_{\mathsf{F}}^{\mathsf{T}} \frac{\partial h_{\mathsf{F}}}{\partial x_{\mathsf{F}}} + h_{\mathsf{F}}^{\mathsf{T}} \frac{\partial f_{\mathsf{F}}}{\partial x_{\mathsf{F}}} = f_{\mathsf{F}}^{\mathsf{T}} h_{\mathsf{F},\mathsf{F}} + h_{\mathsf{F}}^{\mathsf{T}} f_{\mathsf{F},\mathsf{F}}. \tag{A6}$$

The second-derivatives matrix of a product of scalar fns. $f[x_{\mathsf{F}}]$ and $h[x_{\mathsf{F}}]$ is

$$\frac{\partial^2(fh)}{(\partial x_{\mathsf{F}})^2} = \frac{\partial^2 f}{(\partial x_{\mathsf{F}})^2} h + \left(\frac{\partial f}{\partial x_{\mathsf{F}}}\right)^{\mathsf{T}}\left(\frac{\partial h}{\partial x_{\mathsf{F}}}\right) + \left(\left(\frac{\partial f}{\partial x_{\mathsf{F}}}\right)^{\mathsf{T}}\left(\frac{\partial h}{\partial x_{\mathsf{F}}}\right)\right)^{\mathsf{T}} + f \frac{\partial^2 h}{(\partial x_{\mathsf{F}})^2}. \tag{A7}$$

The introduced notation is used now to obtain derivatives of the (*M,m*) Massieu fn. $\Theta^{M,m}[\alpha_{\mathsf{L}}, o_{\mathsf{U}}]$, Eq.(42) or Eq.(52) (remembering that ${}^{\Phi}\Theta^{M,m} = \Theta^{M,m}$), in terms of derivatives of the basic Massieu fn. $\Lambda_{\mathrm{eq}}[\alpha_{\mathsf{F}}]$, Eq.(11) with Eq.(5), or of the effective action fn. $\Phi_{\mathrm{eq}}[o_{\mathsf{F}}]$, Eq.(29). It will be convenient to introduce the notation for the argument of $\Theta^{M,m}$, the vector $x_{\mathsf{F}} = (x_{\mathsf{L}}, x_{\mathsf{U}}) \equiv (\alpha_{\mathsf{L}}, o_{\mathsf{U}}) \in \mathcal{D}^{M,m}$, and for auxiliary vectors: $\tilde{\alpha}_{\mathsf{F}}[x_{\mathsf{F}}] = (x_{\mathsf{L}}, \alpha_{\mathsf{U}}^{M,m}[x_{\mathsf{F}}])$ and $\tilde{o}_{\mathsf{F}}[x_{\mathsf{F}}] = (o_{\mathsf{L}}^{M,m}[x_{\mathsf{F}}], x_{\mathsf{U}})$. This allows to rewrite the mappings (39) and (49) as $x_{\mathsf{F}} \mapsto \tilde{\alpha}_{\mathsf{F}}[x_{\mathsf{F}}] \in \mathcal{A}$, and $x_{\mathsf{F}} \mapsto \tilde{o}_{\mathsf{F}}[x_{\mathsf{F}}] \in \mathcal{O}$. Considering $x_{\mathsf{F}} \in \mathcal{D}^{M,m}$ to be the independent variable vector, we replace $\alpha_{\mathsf{F}}$ and $o_{\mathsf{F}}$ in Eqs.(20) and (24) by $\tilde{\alpha}_{\mathsf{F}}$ and $\tilde{o}_{\mathsf{F}}$ to obtain

$$\Lambda_F\left[\tilde{\alpha}_F\left[x_F\right]\right] = -\tilde{o}_F\left[x_F\right], \qquad \Phi_F\left[\tilde{o}_F\left[x_F\right]\right] = \tilde{\alpha}_F\left[x_F\right]. \tag{A8}$$

In the notation for derivatives of functions $\Lambda_{eq}$ and $\Phi_{eq}$, the subscript 'eq' is suppressed. Using the present notation for Eqs.(42) and (52) we have

$$\Theta^{M,m}\left[x_F\right] = \Lambda_{eq}\left[\tilde{\alpha}_F\left[x_F\right]\right] + x_U^T \tilde{\alpha}_U\left[x_F\right] = \Phi_{eq}\left[\tilde{o}_F\left[x_F\right]\right] - x_L^T \tilde{o}_L\left[x_F\right]. \tag{A9}$$

The gradient of this fn. evaluated with the help of Eqs.(A5) and (A6) is

$$\begin{aligned}
\left(\Theta_F^{M,m}\left[x_F\right]\right)^T &= \left(\Lambda_F\left[\tilde{\alpha}_F\left[x_F\right]\right]\right)^T \tilde{\alpha}_{F,F}\left[x_F\right] + \left(0_L^T \mid \tilde{\alpha}_U^T\left[x_F\right]\right) + x_U^T \tilde{\alpha}_{U,F}\left[x_F\right] \\
&= \left(\Phi_F\left[\tilde{o}_F\left[x_F\right]\right]\right)^T \tilde{o}_{F,F}\left[x_F\right] - \left(\tilde{o}_L^T\left[x_F\right] \mid 0_U^T\right) - x_L^T \tilde{o}_{L,F}\left[x_F\right]
\end{aligned} \tag{A10}$$

($0_L$ and $0_U$ denote zero vectors in appropriate subsets of indices, similarly $0_{UL}$, $0_{UU}$, etc. will denote zero matrices). This gradient is expressed in terms of Jacobian matrices $\tilde{\alpha}_{F,F}$ and $\tilde{o}_{F,F}$. They can be easily evaluated

$$\tilde{\alpha}_{F,F}\left[x_F\right] = \frac{\partial \tilde{\alpha}_F\left[x_F\right]}{\partial x_F} = \begin{pmatrix} 1_{LL} & 0_{UL} \\ \tilde{\alpha}_{U,L} & \tilde{\alpha}_{U,U} \end{pmatrix}, \qquad \tilde{o}_{F,F}\left[x_F\right] = \frac{\partial \tilde{o}_F\left[x_F\right]}{\partial x_F} = \begin{pmatrix} \tilde{o}_{L,L} & \tilde{o}_{U,L} \\ 0_{UL} & 1_{UU} \end{pmatrix} \tag{A11}$$

($1_{LL}$ and $1_{UU}$ denote unit matrices $\{\delta_{ij}\}$ in appropriate subspaces). After inserting the results (A8) and (A11) into the first or the second form of Eq.(A10), the common simple result for $\Theta_F^{M,m}$ follows

$$\Theta_F^{M,m}\left[x_F\right] = \left(-\tilde{o}_L\left[x_F\right], \tilde{\alpha}_U\left[x_F\right]\right) = \left(-o_L^{M,m}\left[\alpha_L, o_U\right], \alpha_U^{M,m}\left[\alpha_L, o_U\right]\right) \tag{A12}$$

(i.e., Eq.(59) is confirmed). Interestingly, this result is independent of the Jacobian matrices.

The second-derivatives matrix, obtained directly from Eq.(A12) by applying Eq.(A3)

$$\Theta_{FF}^{M,m}\left[x_F\right] = \begin{pmatrix} -\tilde{o}_{L,L} & -\tilde{o}_{L,U} \\ \tilde{\alpha}_{U,L} & \tilde{\alpha}_{U,U} \end{pmatrix}, \tag{A13}$$

is composed of blocks of the two Jacobian matrices. The equations that allow their determination are obtained by differentiations in Eq.(A8)

$$\Lambda_{FF} \tilde{\alpha}_{F,F} = -\tilde{o}_{F,F}, \qquad \Phi_{FF} \tilde{o}_{F,F} = \tilde{\alpha}_{F,F}. \tag{A14}$$

This system of equations is solved now for 'nontrivial' blocks of $\tilde{\alpha}_{F,F}$ and $\tilde{o}_{F,F}$, Eq.(A11). Presence of the trivial blocks (zero and unit) is helpful, it leads to the following equations for blocks

$$\Lambda_{UL} + \Lambda_{UU}\tilde{\alpha}_{U,L} = 0_{UL}, \qquad \Lambda_{UU}\tilde{\alpha}_{U,U} = -1_{UU}, \tag{A15}$$

$$\Phi_{LL}\tilde{o}_{L,L} = 1_{LL}, \qquad \Phi_{LL}\tilde{o}_{L,U} + \Phi_{LU} = 0_{LU}, \tag{A16}$$

which have the solutions

$$\tilde{\alpha}_{U,L} = -\Lambda_{UU}^{-1}\Lambda_{UL}, \qquad \tilde{\alpha}_{U,U} = -\Lambda_{UU}^{-1}, \tag{A17}$$

$$\tilde{o}_{L,L} = \Phi_{LL}^{-1}, \qquad \tilde{o}_{L,U} = -\Phi_{LL}^{-1}\Phi_{LU}. \tag{A18}$$

After inserting these solutions into remaining equations for blocks, one obtains the second form of solutions

$$\tilde{\alpha}_{U,L} = \Phi_{UL}\Phi_{LL}^{-1}, \qquad \tilde{\alpha}_{U,U} = \Phi_{UU} - \Phi_{UL}\Phi_{LL}^{-1}\Phi_{LU}, \tag{A19}$$

$$\tilde{o}_{L,L} = -\Lambda_{LL} + \Lambda_{LU}\Lambda_{UU}^{-1}\Lambda_{UL}, \qquad \tilde{o}_{L,U} = \Lambda_{LU}\Lambda_{UU}^{-1}. \tag{A20}$$

Since, as shown in Sec.II, $\Lambda_{FF}$ is positive definite and $\Phi_{FF}$ is negative definite, hence, as discussed in Appendix B, their diagonal blocks $\Lambda_{UU}$ and $\Phi_{LL}$ preserve definiteness and, therefore, their reciprocals $\Lambda_{UU}^{-1}$ and $\Phi_{LL}^{-1}$ exist and have the definiteness unchanged. By inserting the results (A17)–(A20) into Eq.(A13), two forms of $\Theta_{FF}^{M,m}$ in Eq.(64) are obtained.

The derivatives of the Gibbs-Helmholtz fn. $\Upsilon^{M,m}$, Eq.(66), are obtained in a similar way as the derivatives of $\Theta^{M,m}$. The vector of independent variables of $\Upsilon^{M,m}$ is denoted $z_F = (z_1, z_{L'}, z_U) \equiv (\beta, \gamma_{L'}, o_U)$ and depending on it the vector $y_F[z_F] = (\hat{\alpha}_L[\beta, \gamma_{L'}], o_U)$ $= (\beta, -\beta\gamma_{L'}, o_U)$, [see Eq.(67)] denotes the argument of $\Theta^{M,m}$ in Eq.(66). Using the present notation, Eq.(66) can be rewritten as

$$\Upsilon^{M,m}[z_F] = -z_1^{-1}\Theta^{M,m}[y_F[z_F]] \tag{A21}$$

The first-derivative vector, according to transposed Eq.(A5) is

$$\Upsilon_F^{M,m}[z_F] = -\Theta^{M,m}\left(\frac{\partial z_1^{-1}}{\partial z_F}\right)^T - z_1^{-1}\left(\frac{\partial \Theta^{M,m}[y_F[z_F]]}{\partial z_F}\right)^T \quad (A22)$$

$$= z_1^{-2}\Theta^{M,m}[y_F[z_F]](1, 0_{F'}) - z_1^{-1}(y_{F,F}[z_F])^T \Theta_F^{M,m}[y_F[z_F]],$$

where the Jacobian matrix is

$$y_{F,F}[z_F] = \begin{pmatrix} 1 & 0_{L'}^T & 0_{LU} \\ -z_{L'} & -z_1 1_{L'L'} & \\ 0_{UL} & & 1_{UU} \end{pmatrix}. \quad (A23)$$

After inserting it into Eq.(A22) we find

$$\Upsilon_F^{M,m}[z_F] = \left(z_1^{-2}\Theta^{M,m} - z_1^{-1}\Theta_1^{M,m} + z_1^{-1}z_{L'}^T\Theta_{L'}^{M,m}, \Theta_{L'}^{M,m}, -z_1^{-1}\Theta_U^{M,m}\right)$$

$$= \left(\beta^{-2}\Theta^{M,m} + \beta^{-1}\left(o_1^{M,m} - \gamma_{L'}^T o_{L'}^{M,m}\right), -o_{L'}^{M,m}, -\beta^{-1}\alpha_U^{M,m}\right), \quad (A24)$$

where, in the last step we took $\Theta_F^{M,m}$ from Eq.(A12). After taking for $-\beta^{-1}\Theta^{M,m} = \Upsilon^{M,m}$ the third line of Eq.(68), we transform finally Eq.(A24) into the result (70).

The second-derivatives matrix obtained from Eq.(A21) according to (A7) is

$$\Upsilon_{FF}^{M,m}[z_F] = \Upsilon_{FF}^{(a)} + \Upsilon_{FF}^{(b)} + \Upsilon_{FF}^{(c)} + \Upsilon_{FF}^{(d)} \quad (A25)$$

with

$$\Upsilon_{FF}^{(a)} = -\frac{\partial^2(z_1^{-1})}{(\partial z_F)^2}\Theta^{M,m} = -2z_1^{-3}\Theta^{M,m}\begin{pmatrix} 1 & 0_{F'}^T \\ 0_{F'} & 0_{F'F'} \end{pmatrix}, \quad (A26)$$

$$\Upsilon_{FF}^{(b)} = -\left(\frac{\partial(z_1^{-1})}{\partial z_F}\right)^T \frac{\partial \Theta^{M,m}[y_F]}{\partial z_F} = z_1^{-2}(1, 0_{F'})(\Theta_F^{M,m})^T y_{F,F} = \begin{pmatrix} z_1^{-2}(\Theta_F^{M,m})^T y_{F,F} \\ 0_{F'F} \end{pmatrix}, \quad (A27)$$

$$\Upsilon_{FF}^{(c)} = \left(\Upsilon_{FF}^{(b)}\right)^T, \quad (A28)$$

$$\Upsilon_{FF}^{(d)} = -z_1^{-1}\frac{\partial^2(\Theta^{M,m}[y_F[z_F]])}{(\partial z_F)^2} = -z_1^{-1}\frac{\partial}{\partial z_F}\left((y_{F,F}[z_F])^T \Theta_F^{M,m}[y_F[z_F]]\right)$$

$$= \Upsilon_{FF}^{(d1)} + \Upsilon_{FF}^{(d2)} \quad (A29)$$

where, as it can be easily verified via components of matrices and vectors,

$$\Upsilon_{FF}^{(d1)} = -z_1^{-1} \left(\Theta_F^{M,m}\right)^T \frac{\partial y_{F,F}[z_F]}{\partial z_F} = -z_1^{-1} \left(\Theta_F^{M,m}\right)^T y_{F,FF}, \tag{A30}$$

$$\Upsilon_{FF}^{(d2)} = -z_1^{-1} y_{F,F}^T \frac{\partial \Theta_F^{M,m}[y_F[z_F]]}{\partial z_F} = -z_1^{-1} y_{F,F}^T \Theta_{FF}^{M,m} y_{F,F}. \tag{A31}$$

Since the non-constant blocks of $y_{F,F}$, Eq.(A23), are only $y_{L',1} = -z_{L'}$ and $y_{L',L'} = -z_1 1_{L'L'}$, the non-zero blocks of $y_{F,FF}$ are $y_{L',1L'} = -1_{L'L'}$ and $y_{L',L'1} = -1_{L'L'}$. This allows evaluation of Eq.(A30) as

$$\Upsilon_{FF}^{(d1)} = \begin{pmatrix} 0 & z_1^{-1}\left(\Theta_{L'}^{M,m}\right)^T & 0_U^T \\ \hline z_1^{-1}\Theta_{L'}^{(M,m)} & & \\ 0_U & & 0_{F'F'} \end{pmatrix}. \tag{A32}$$

For evaluation of $\Upsilon_{FF}^{(b)}$ we need $\partial \Theta^{(M,m)}[y_F]/\partial z_F$, which can be obtained with the help of Eqs. (A5) and (A23) as

$$\left(\Theta_F^{M,m}\right)^T y_{F,F} = \left(\Theta_1^{M,m} - \left(\Theta_{L'}^{M,m}\right)^T z_{L'} \mid -z_1\left(\Theta_{L'}^{M,m}\right)^T \mid \left(\Theta_U^{M,m}\right)^T\right). \tag{A33}$$

Finally

$$\Upsilon_{FF}^{M,m}[z_F] = \Upsilon_{FF}^{(ad)} + \Upsilon_{FF}^{(d2)}, \tag{A34}$$

where

$$\Upsilon_{FF}^{(ad)}[z_F] = \Upsilon_{FF}^{(a)} + \Upsilon_{FF}^{(b)} + \Upsilon_{FF}^{(c)} + \Upsilon_{FF}^{(d1)}$$
$$= \begin{pmatrix} -2z_1^{-3}\Theta^{M,m} + 2z_1^{-2}\left(\Theta_1^{M,m} - z_{L'}^T\Theta_{L'}^{M,m}\right) & 0_{L'}^T & z_1^{-2}\left(\Theta_U^{M,m}\right)^T \\ \hline 0_{L'} & & \\ z_1^{-2}\Theta_U^{M,m} & & 0_{F'F'} \end{pmatrix}. \tag{A35}$$

Vanishing of blocks $1L'$ and $L'1$ is to be noted. Eq.(A35) can be rewritten as

$$\Upsilon_{FF}^{(ad)} = \begin{pmatrix} \Upsilon_{11}^{(ad)} & 0_{L'}^T & \left(\Upsilon_{U1}^{(ad)}\right)^T \\ 0_{L'} & & \\ \Upsilon_{U1}^{(ad)} & & 0_{F'F'} \end{pmatrix} \quad (A36)$$

where

$$\Upsilon_{11}^{(ad)} = -2\beta^{-3} S^{BGS}\left[\hat{\Gamma}_{eq}^{M,m}\right], \quad (A37)$$

$$\Upsilon_{U1}^{(ad)} = -\beta^{-1}\gamma_U^{M,m} \quad (A38)$$

Eqs.(59), (66) and (68) were helpful in obtaining the final expressions. The result Eq.(71) corresponds to (A34) with Eqs. (A36)–(A38) and (A31).

# APPENDIX B: PROPERTIES OF A POSITIVE (NEGATIVE) DEFINITE MATRIX

For completeness of our paper, we recall shortly well known facts[76] about definiteness of a real symmetric matrix $f_{FF}$. It is called *positive definite* if

$$x_F^T f_{FF} x_F > 0 \quad (B1)$$

for all nonzero real vectors $x_F$. This property holds if and only if there exists a real matrix $g_{FF}$, $\det(g_{FF}) \neq 0$, such that

$$f_{FF} = g_{FF} g_{FF}^T. \quad (B2)$$

Another necessary and sufficient condition for a real symmetric matrix to be positive definite is that all its eigenvalues are positive. The matrix $f_{FF}^{-1}$ — the inverse of a positive definite matrix $f_{FF}$ is also positive definite. The definition of positive definiteness is equivalent to the requirement that the determinants associated with all upper-left square submatrices are positive.

Any diagonal block of a positive definite matrix is also positive definite. To verify this, the full set of indices is arbitrarily partitioned into three subsets: $F = A \cup B \cup C$ (any one of $A$, $B$, $C$ can be empty set). For the proof we chose the following vector $x_F^T = \left(0_A^T \mid x_B^T \mid 0_C^T\right)$, where $x_B$ is a nonzero vector, while $0_A$ and $0_C$ denote zero vectors within the index subsets $A$ and $C$. From the assumed inequality (B1) follows

$$0 < \left(0_A^T \mid x_B^T \mid 0_C^T\right) \begin{pmatrix} f_{AA} & f_{AB} & f_{AC} \\ f_{BA} & f_{BB} & f_{BC} \\ f_{CA} & f_{CB} & f_{CC} \end{pmatrix} \begin{pmatrix} 0_A \\ x_B \\ 0_C \end{pmatrix} = x_B^T f_{BB} x_B, \quad (B3)$$

thus the diagonal block $f_{BB}$ is positive definite.

When the relation '>' is replaced by '≥' in Eq.(B1) the matrix $f_{FF}$ is called *positive semidefinite*. A real symmetric matrix $f_{FF}$ is *negative* definite (semidefinite) if $(-f_{FF})$ is positive definite (semidefinite). Its properties follow directly from the last replacement.

If a scalar real function $f$ of a vector $x_F$ is twice-differentiable then: (i) from positive definitness of its second derivative matrix $f_{FF}(x_F)$ [see Eq.(A3)] follows strict convexity of $f(x_F)$ [see Eq.(22) for this property of $\Lambda_{eq}$ as a function of $\alpha_F$], (ii) from strict convexity of $f(x_F)$ follows positive semidefinitness of $f_{FF}(x_F)$.

## APPENDIX C: LEGENDRE TRANSFORMATION

Since many thermodynamic fns. are defined in the present paper via Legendre transformation of some basic fn., we recall shortly the definition of this transformation[77] and discuss its basic properties. Given a scalar function $f(x_A, y_B)$ of two vector variables $x_A$, $y_B$ (their dimensionalities can be different). The *Legendre transform* $g(u_A, y_B)$ of $f(x_A, y_B)$, introduced in connection with the change of variables — the mapping $(x_A, y_B) \mapsto (u_A, y_B)$, which is induced by the "generating property" of $f(x_A, y_B)$

$$\left( \frac{\partial f(x_A, y_B)}{\partial x_A} \right)^T = u_A, \tag{C1}$$

is defined by

$$g(u_A, y_B) = f(x_A, y_B) - u_A^T x_A. \tag{C2}$$

When the variables $(u_A, y_B)$ are taken as independent, the vector $x_A$ on the right-hand side of Eq.(C2) is a fn. of $(u_A, y_B)$. Stemming from the reverse mapping (assumed to exist) $(u_A, y_B) \mapsto (x_A, y_B)$, this fn. is the solution of Eq.(C1) w.r.t. $x_A$ at given $(u_A, y_B)$.

The Legendre transform $g(u_A, y_B)$ shows the following generating property

$$\left( \frac{\partial g(u_A, y_B)}{\partial u_A} \right)^T = -x_A, \tag{C3}$$

while the derivative w.r.t. $y_B$ is common for $g$ and $f$:

$$\left( \frac{\partial g(u_A, y_B)}{\partial y_B} \right) = \left( \frac{\partial f(x_A, y_B)}{\partial y_B} \right) \tag{C4}$$

provided $u_A$ and $x_A$ are related as in Eq.(C1) or (C3).

Using the above definitions, the Legendre transform $h(x_A, y_B)$ of $g(u_A, y_B)$, induced by the generating property (C3), is found to be

$$h(x_A, y_B) = g(u_A, y_B) - (-x_A)^T u_A. \tag{C5}$$

From comparison of Eqs. (C5) and (C2) follows immediately that $h(x_A, y_B)$ is identical with $f(x_A, y_B)$. Thus the functions $f(x_A, y_B)$ and $g(u_A, y_B)$, showing the generating properties (C1) and (C3) are *Legendre-transformation equivalent*.

Using for the first derivatives vectors the notation of Appendix A, like $\left(\partial f(x_A, y_B)/\partial x_A\right)^T = f_A(x_A, y_B)$, we summarize mappings connected with the considered Legendre transformations as

$$(x_A, y_B) \mapsto (u_A, y_B) = (f_A(x_A, y_B), y_B), \tag{C6}$$

$$(u_A, y_B) \mapsto (x_A, y_B) = (-g_A(u_A, y_B), y_B). \tag{C7}$$

Their superposition leads to identities

$$f_A(-g_A(u_A, y_B), y_B) = u_A \qquad \forall (u_A, y_B), \tag{C8}$$

$$-g_A(f_A(x_A, y_B), y_B) = x_A \qquad \forall (x_A, y_B). \tag{C9}$$

Eq.(C4) can be rewritten in two explicit versions:

$$g_B(u_A, y_B) = f_B(-g_A(u_A, y_B), y_B) \qquad \forall (u_A, y_B), \tag{C10}$$

$$f_B(x_A, y_B) = g_B(f_A(x_A, y_B), y_B) \qquad \forall (x_A, y_B). \tag{C11}$$

## APPENDIX D: IDENTITIES DUE TO VARIABLES TRANSFORMATIONS

Various Massieu fns. defined in Sec. II and Sec. III, are mutually related with appropriate Legendre transformations (see Table I). The accompanying mappings (transformations) $\mathcal{T}_x^y$ of the independent-variables vectors are displayed schematically in Table II. The subscript $x$ of $\mathcal{T}_x^y$ characterizes the domain of the map, while the superscript $y$ — the range (codomain) of the map. As shown in Table II, $x$ or $y$ may mean '$M,M$', '$M,0$', '$M,m$'. Various variables domains $\mathcal{D}^z$ are related by the same maps

$$\mathcal{D}^x \ni \mathcal{X} \mapsto \mathcal{T}_x^y[\mathcal{X}] = \mathcal{Y} \in \mathcal{D}^y. \tag{D1}$$

Here $\mathcal{X}$ or $\mathcal{Y}$ may mean anyone of three vectors: $\alpha_F \equiv (\alpha_L, \alpha_U)$, $o_F \equiv (o_L, o_U)$, $(\alpha_L, o_U)$. We observe in Table 2 that domain points may be connected by various paths (series of mappings), e.g., $\mathcal{T}_y^x[\mathcal{T}_x^y[\mathcal{X}]] = \mathcal{X}$, $\mathcal{T}_y^x[\mathcal{Y}] = \mathcal{T}_z^x[\mathcal{T}_y^z[\mathcal{Y}]]$. This leads to specific identities satisfied by four vector functions that are involved in mappings: $o_F^{eq}[\alpha_F]$ in $\mathcal{T}_{M,M}^{M,0}$, Eq.(21), and $\mathcal{T}_{M,M}^{M,m}$, Eq.(35), $\alpha_F^{eq}[o_F]$ in $\mathcal{T}_{M,0}^{M,M}$, Eq.(24), and $\mathcal{T}_{M,0}^{M,M}$, Eq.(46), $\alpha_U^{M,m}[\alpha_L, o_U]$ in $\mathcal{T}_{M,m}^{M,M}$, Eq.(39), and $o_L^{M,m}[\alpha_L, o_U]$ in $\mathcal{T}_{M,m}^{M,0}$, Eq.(49). Obtained earlier two identities, Eqs. (26), (27), correspond to return paths between $\mathcal{O} \equiv \mathcal{D}^{M,0}$ and $\mathcal{A} \equiv \mathcal{D}^{M,M}$, i.e., $\mathcal{T}_{M,M}^{M,0}[\mathcal{T}_{M,0}^{M,M}[o_F]] = o_F$, and between $\mathcal{A}$ and $\mathcal{O}$. The return paths between $\mathcal{A}$ and $\mathcal{D}^{M,m}$ and vice versa result in

$$\alpha_U^{M,m}[\alpha_L, o_U^{eq}[\alpha_F]] = \alpha_U \qquad \forall \alpha_F \in \mathcal{A}, \tag{D2}$$

$$o_U^{eq}[\alpha_L, \alpha_U^{M,m}[\alpha_L, o_U]] = o_U \qquad \forall (\alpha_L, o_U) \in \mathcal{D}^{M,m}, \tag{D3}$$

while the same involving $\mathcal{O}$ and $\mathcal{D}^{M,m}$ give

$$o_L^{M,m}[\alpha_L^{eq}[o_F], o_U] = o_L \qquad \forall o_F \in \mathcal{O}, \tag{D4}$$

$$\alpha_L^{eq}[o_L^{M,m}[\alpha_L, o_U], o_U] = \alpha_L \qquad \forall (\alpha_L, o_U) \in \mathcal{D}^{M,m}. \tag{D5}$$

Similarly, one can obtain identities corresponding to longer paths. However, they all happen to be equivalent to already knows identities stemming from the Legendre transformations connecting $\mathcal{A}$ with $\mathcal{D}^{M,m}$ and $\mathcal{O}$ with $\mathcal{D}^{M,m}$, namely Eqs.(44), (45), (54), (55). We rewrite them here in terms of the mapping functions with the help of Eqs.(36), (40), (47) and (50), and remembering that $^\Phi\Theta^{M,m} \equiv \Theta^{M,m}$:

$$o_L^{M,m}[\alpha_L, o_U] = o_L^{eq}[\alpha_L, \alpha_U^{M,m}[\alpha_L, o_U]] \qquad \forall (\alpha_L, o_U) \in \mathcal{D}^{M,m}, \tag{D6}$$

$$o_L^{M,m}[\alpha_L, o_U^{eq}[\alpha_F]] = o_L^{eq}[\alpha_F] \qquad \forall \alpha_F \in \mathcal{A}, \tag{D7}$$

$$\alpha_U^{M,m}\left[\alpha_L, o_U\right] = \alpha_U^{eq}\left[o_L^{M,m}\left[\alpha_L, o_U\right], o_U\right] \qquad \forall \left(\alpha_L, o_U\right) \in \mathcal{D}^{M,m}, \tag{D8}$$

$$\alpha_U^{M,m}\left[\alpha_L^{eq}\left[o_F\right], o_U\right] = \alpha_U^{eq}\left[o_F\right] \qquad \forall o_F \in \mathcal{O}. \tag{D9}$$

Since the Massieu fns. $\Lambda_{eq}$, $\Phi_{eq}$, $\Theta^{M,m}$ are more fundamental characteristics of various systems than the mapping fns., we rewrite all identities expressing the mapping fns. in terms of the first-derivatives vectors $\Lambda_F$, $\Phi_F$, $\Theta_F^{M,m}$, Eqs.(21), (24), (50), (40). They are given as second versions of Eqs.(26), (27), and some equations from the set (44)-(55)

$$\Theta_U^{M,m}\left[\alpha_L, -\Lambda_U\left[\alpha_F\right]\right] = \alpha_U \qquad \forall \alpha_F \in \mathcal{A}, \tag{D10}$$

$$\Lambda_U\left[\alpha_L, \Theta_U^{M,m}\left[\alpha_L, o_U\right]\right] = -o_U \quad \forall \left(\alpha_L, o_U\right) \in \mathcal{D}^{M,m}, \tag{D11}$$

$$\Theta_L^{M,m}\left[\Phi_L\left[o_F\right], o_U\right] = -o_L \qquad \forall o_F \in \mathcal{O}, \tag{D12}$$

$$\Phi_L\left[-\Theta_L^{M,m}\left[\alpha_L, o_U\right], o_U\right] = \alpha_L \quad \forall \left(\alpha_L, o_L\right) \in \mathcal{D}^{M,m}, \tag{D13}$$

$$\Theta_F^{M,m}\left[\alpha_L, o_U\right] = \left(\Lambda_L\left[\alpha_L, \Theta_U^{M,m}\left[\alpha_L, o_U\right]\right], \Phi_U\left[-\Theta_L^{M,m}\left[\alpha_L, o_U\right], o_U\right]\right) \quad \forall \left(\alpha_L, o_U\right) \in \mathcal{D}^{M,m}, \tag{D14}$$

$$\Theta_L^{M,m}\left[\alpha_L, -\Lambda_U\left[\alpha_F\right]\right] = \Lambda_L\left[\alpha_F\right] \qquad \forall \alpha_F \in \mathcal{A}, \tag{D15}$$

$$\Theta_U^{M,m}\left[\Phi_L\left[o_F\right], o_U\right] = \Phi_U\left[o_F\right] \qquad \forall o_F \in \mathcal{O}. \tag{D16}$$

Note that Eqs.(D10)–(D13) correspond to Eqs.(D2)–(D5); Eq.(D14) – to Eqs.(D6), (D8); Eqs. (D15), (D16) – to Eqs.(D7), (D9).

**Table I.** Thermodynamic state functions and their relations.

| for fully open system | for fully closed system |
|---|---|
| $\Lambda_{eq}[\alpha_F] \equiv \Theta^{M,M}[\alpha_F]$ | $\Phi_{eq}[o_F] = S^{BGS}[\hat{\Gamma}_{eq}^{ev}[o_F]] \equiv \Theta^{M,0}[o_F]$ |
| the basic Massieu fn. | the effective action fn. = the entropy |

<table>
<tr><td colspan="2" align="center">

for partially open/closed system

$\Theta^{M,m}[\alpha_L, o_U]$

the $(M,m)$ Massieu fn., $0 \leq m \leq M$

- - - - - - - - - - - - - - - - - - - - - - - - - - - - - - - - - - - - -

$\Upsilon^{M,m}[\beta, \gamma_{L'}, o_U]$

the $(M,m)$ Gibbs-Helmholtz fn., $1 \leq m \leq M$

</td></tr>
</table>

Massieu fns.

$\Lambda_{eq}[\alpha_F] \equiv \Theta^{M,M}[\alpha_F]$ $\xrightleftharpoons{\text{Legendre transformations w.r.t. all variables}}$ $\Theta^{M,0}[o_F] \equiv \Phi_{eq}[o_F]$

Legendre transformations w.r.t. some variables

$\Theta^{M,m}[\alpha_L, o_U]$

Massieu-Planck transformation
direct ↓↑ reverse

Gibbs-Helmholtz fns. $\Upsilon^{M,m}[\beta, \gamma_{L'}, o_U]$

**Table II**. Variables transformations.

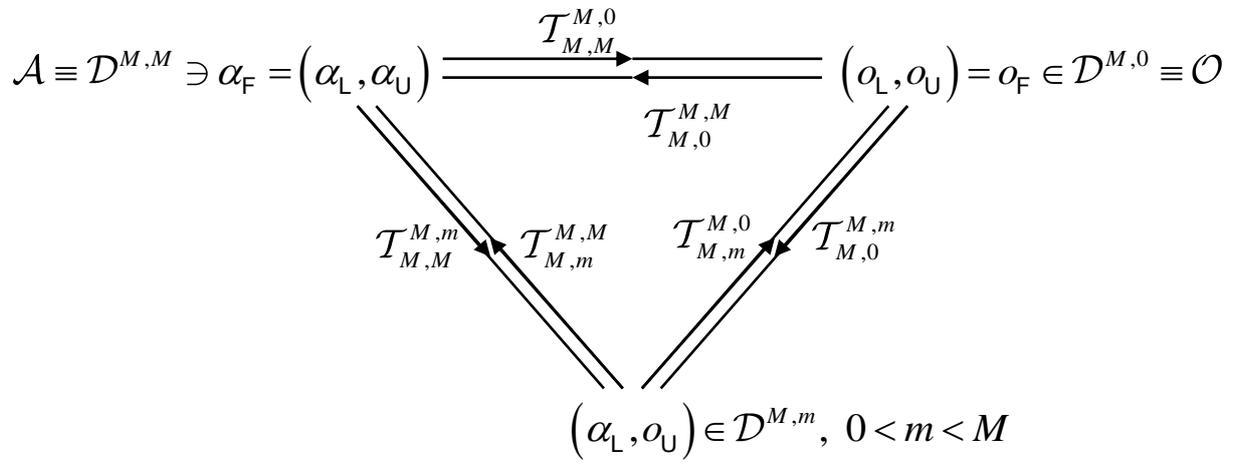